\documentclass[twocolumn]{aastex631}

\usepackage{amsmath}

\newcommand{\eq}[1]{Eq.~\eqref{#1}}
\newcommand{\fref}[1]{Fig.~\ref{#1}}
\newcommand{\sref}[1]{Sec.~\ref{#1}}
\newcommand{\tref}[1]{Table~\ref{#1}}

\newcommand{\dd}{\mathrm{d}}

\begin{document}

\title{Detection of Intermediate-Mass Ratio Inspirals in Globular Clusters: Revealing the Brownian Motion with Gravitational Waves}

\author[0000-0002-5467-3505]{Alejandro Torres-Orjuela}
\email{Corresponding author: atorreso@bimsa.cn}
\affiliation{Beijing Institute of Mathematical Sciences and Applications, Beijing 101408, China}

\author[0000-0002-9458-8815]{Ver\'{o}nica V\'{a}zquez-Aceves}
\affiliation{Kavli Institute for Astronomy and Astrophysics at Peking University, 100871 Beijing, China}

\author[0009-0006-3173-6506]{Tian-Xiao Wang}
\affiliation{MOE Key Laboratory of TianQin Mission, TianQin Research Center for Gravitational Physics $\&$ Frontiers Science Center for TianQin, Gravitational Wave Research Center of CNSA, Sun Yat-sen University (Zhuhai Campus), Zhuhai 519082, China}



\begin{abstract}

Intermediate-mass ratio inspirals (IMRIs) formed by stellar-mass compact objects orbiting intermediate-mass black holes will be detected by future gravitational wave (GW) observatories like TianQin, LISA, and AION. We study a set of 100 IMRI systems in globular clusters obtained from MOCCA simulations to estimate their detectability. Furthermore, we model the Brownian motion of the IMRIs induced by weak interactions with the surrounding field of stars and include its effect on the GW's phase through Doppler and aberrational phase shift. We find that a small fraction of IMRIs ($<10\,\%$) will have signal-to-noise ratios (SNR) high enough to be detected by TianQin, LISA, and AION. However, for all sources detected, the SNR is high enough to discern the Brownian motion of the IMRI. More precisely, we find that the match between the signal containing the effect of the Brownian motion and a waveform model without this effect is mostly low ($<0.8$). These results highlight the importance of including the interaction of IMRIs with the surrounding field of stars to obtain proper detection, but also show the possibility of studying the environment of the source using GWs.

\end{abstract}

\keywords{Gravitational wave sources(677) --- Intermediate-mass black holes(816) --- Gravitational interaction(669) --- Star clusters(1567) --- Gravitational wave astronomy(675)}


\section{Introduction} \label{sec:intro}

Gravitational wave (GW) astronomy is a vibrant field with numerous detections in the high-frequency band above $10\,{\rm Hz}$ covered by LIGO, Virgo, and KAGRA detectors~\cite{GWTC1,GWTC2,GWTC3} and the detection of a signal in the $\rm nHz$-band covered by pulsar timing arrays~\citep{ipta_2013,nanograv_2013,ppta_2013,cpta_2016,epta_2016,inpta_2018}. In coming years, further detection windows will be opened by the addition of space-based laser interferometry detectors like TianQin, LISA, and Taiji~\citep{tianqin_2021,lisa_2024,taiji_2015} covering the low-frequency band ($\rm mHz$) as well as ground-based and space-based atom interferometry detectors such as AION, ZAIGA, and AEDGE~\citep{aion_2020,zaiga_2020,aedge_2020} or space-based laser interferometry detectors such as DECIGO~\citep{decigo_2021} sensitive in the intermediate band ($\rm dHz$).

These upcoming detectors will allow us to study a variety of new sources, including intermediate-mass ratio inspirals (IMRIs) formed by a stellar-mass compact object inspiraling into an intermediate-mass black hole (IMBH) of $10^{2}-10^{5} \ \rm M_{\odot}$~\citep{lisa_2022a,tianqin_2024,abend_allard_2023,abdalla_abe_2024}. IMRIs, with mass ratios between $10^{-1}$ to $10^{-4}$, can form in dense stellar systems like Globular Clusters (GCs) or Nuclear Star Clusters (NSC); however, they rely on the existence of IMBHs. The detection of a binary merger that resulted in a $142\,{\rm M_{\odot}}$ black hole (BH) by the LIGO/Virgo Collaboration in 2020~\citep{ligo_virgo_2020e} indicates the existence of IMBHs in the light-mass end. From electromagnetic observations, there is evidence that the stellar cluster G1 harbors an IMBH with a mass of approximately $2\times10^4\,{\rm M_{\odot}}$~\citep{gebhardt_rich_2002}. Similarly, an IMBH with a mass of $\approx1.7\times10^4\,{\rm M_{\odot}}$ has been proposed to reside in the GC NGC 6388~\citep{lutzgendorf_kissler-patig_2011}. Omega Centauri also appears to host an IMBH, with estimates ranging from $10^4$ to $10^5\,{\rm M_{\odot}}$~\citep{noyola_gebhardt_2008,noyola_gebhardt_2010,miocchi_2010,jalali_baumgardt_2012} while the dwarf galaxy RGG 118 is suspected to host an IMBH of around $5\times10^4\,{\rm M_\odot}$~\citep{baldassare_reines_2015}. However, the evidence for these systems remains controversial.

The formation of IMBHs and IMRIs appears in different numerical simulations of NSC and GC through different channels, e.g., repeated mergers resulting from binary-single or binary-binary interactions or dynamical captures~\citep{konstantinidis_amaro-seoane_2013,giersz_leigh_2015,rizzuto_naab_2021}. These studies allow us to understand the characteristics of these systems as well as their detectability with GWs, despite the absence of direct detection through electromagnetic observations. The dense environments of NSCs and GCs, where core densities can reach $\gtrsim10^6-10^7\,{\rm M_{\odot}\,pc^{-3}}$, create ideal conditions for the formation of IMBHs and IMRIs. Moreover, other environments like merging galaxies~\citep{vazquez-aceves_amaro-seoane_2023} and AGN disks~\citep{peng_chen_2023}, have also been proposed as potential IMBH and IMRI host systems.

In this work, we focus on GCs as host systems. The high densities in their cores make them very dynamically active systems where IMRIs can form and experience frequent weak gravitational encounters during their evolution. This effect, known as Brownian Motion~\citep{einstein_1905,smoluchowski_1906}, differs from the two-body relaxation processes, although these two processes involve random interactions and lead to gradual changes over time. Two-body relaxation processes redistribute energy and angular momentum, leading to changes in the dynamical structure of the system, causing mass segregation and core collapse in a GC~\citep{binney_tremaine_2008}. The Brownian motion does not change the orbital parameters of the binary but due to the relatively low mass of the IMRIs, $10^{2}-10^{5}\,{\rm M_{\odot}}$, these gravitational encounters can significantly influence the velocity of the IMRI~\citep{merritt_2001,merritt_2002,chatterjee_hernquist_2002,merritt_berczik_2007,bortolas_gualandris_2016,lingam_2018,di-cintio_ciotti_2020,roupas_2021} and, consequently, its GW signal. In contrast, heavier BHs, such as supermassive BHs (SMBHs), are far less susceptible to the motion induced by gravitational interactions, and even with frequent encounters, the Brownian Motion might be negligible. For IMRIs formed in a nuclear star cluster or AGN disc, other factors, such as the potential of the central SMBH and the presence of gas, must be considered; these factors will likely dominate the dynamics, making the effect of the Brownian Motion less important. 

In this paper, we explore whether the Brownian motion of an IMRI in a GC can be detected using the GWs emitted. Multiple effects can occur when a binary emitting GWs interacts with a third body -- depending on the magnitude and duration of the interaction. These effects are, e.g., causing a misalignment of the BHs' spins~\citep{antonini_rodriguez_2018,liu_lai_2019,yu_ma_2020}, changing the binaries eccentricity~\citep{wen_2003,antonini_murray_2014,hoang_naoz_2018,martinez_fragione_2020,dallamico_mapelli_2024}, causing a hardening of the binary or disrupting it~\citep{mandel_brown_2008,trani_spera_2019,marin-pina_gieles_2025}, or inducing a phase shift of the GWs emitted~\citep{torres-orjuela_chen_2020,toubiana_sberna_2021,sberna_babak_2022,hendriks_zwick_2024,zwick_tiede_2024}. We focus on weak and relatively short interactions that do not significantly alter the internal properties of the IMRI. The main observable is then the Doppler and aberrational phase shift induced by an acceleration of the source's center-of-mass~\citep{torres-orjuela_chen_2021}.

In \sref{sec:brown} we derive a simple model of the Brownian motion in a GC to obtain the magnitude of the acceleration, the duration of an encounter, and the frequency of encounters. We introduce in \sref{sec:dat} basic concepts of GW data analysis relevant to our study and in \sref{sec:eff}, we discuss the effect the Brownian motion has on the GWs emitted by an IMRI. In \sref{sec:det}, we consider a set of 100 IMRIs in GCs obtained from the Database Survey I of MOCCA simulations~\citep{askar_szkudlarek_2016} to analyze their detection by TianQin, LISA, and AION. Subsequently, we study the detectability of the Brownian motion from the GW phase shift induced by the encounters. We conclude our paper with a discussion of our results and future prospects in \sref{sec:con}.

\section{Brownian motion in a star cluster}\label{sec:brown}

The Brownian motion of an IMRI in a star cluster is mainly dictated by the properties of the star cluster and the subsequent encounter between the components of the star cluster and the IMRI. Therefore, we model the star density $\eta$ in the cluster to be proportional to the mass density of a simple Plummer model~\citep{binney_tremaine_2008}
\begin{equation}\label{eq:stardens}
    \eta(r) = \frac{3}{4\pi b^3}\left(1+\frac{r^2}{b^2}\right)^{-5/2},
\end{equation}
where $b$ is the Plummer scale length, and $r$ is the distance from the center of the system. A Plummer model is an idealized representation of real star clusters but has the benefit that analytic solutions of different properties can be obtained. 

The star density falls off as the radius $r$ increases, but only becomes zero at infinity. However, real star clusters are not infinitely big, and we need a way to define their size. We use the half-mass radius $R_h$ which corresponds to the radius where half of the mass in the cluster $M$ is contained~\citep{binney_tremaine_2008}. $R_h$ is then defined as the radius where the integral of $M\eta(r)$ in spherical coordinates equals to $M/2$. We find
\begin{equation}\label{eq:size}
    R_h = \left(4^{1/3}-1\right)^{-1/2}b \approx \frac43b,
\end{equation}
which allows us to link the Plummer scale length to the size of the cluster.

\subsection{IMRI orbital parameters and host systems}\label{sec:parasys}

For our analysis of the effect of the Brownian motion on IMRIs, we want to consider realistic data from binary systems that evolved in GCs; nevertheless, this information is only available through numerical simulations. 
We take data from the MOCCA-SURVEY Database I, which consists of about 2000 simulations of GCs~\citep{askar_szkudlarek_2016} dynamically evolved with the MOCCA (MOnte Carlo Cluster simulAtor) code~\citep{hypki_giersz_2012,giersz_heggie_2013}. 

The MOCCA code is based on Michel Hénon's Monte Carlo method~\citep{henon_1971a,henon_1971b}, improved later by \cite{stodolkiewicz_1986} and \cite{giersz_2001,giersz_2001b,giersz_spurzem_2003}. It uses a statistical treatment of two-body relaxation processes to dynamically evolve spherically symmetric stellar clusters. The code also includes stellar and binary evolution prescriptions from the SSE~\citep{hurley_pols_2000} and BSE~\citep{hurley_tout_2002} codes and the N-body treatment FEWBODY~\citep{fregeau_cheung_2004} to handle binary-single and binary-binary encounters. These are important encounters that lead to the formation and growth of IMBHs and the formation of IMRIs. 

Although full N-body simulations of dense stellar systems provide the most accurate and realistic results, the computational cost is extremely high; for example, the NBODY6++GPU code can simulate a system with a million particles producing results within a year~\citep{wang_spurzem_2015, wang_spurzem_2016}, but with current technology, the simulation of thousands of stellar systems via full N-body simulations is not possible. In the case of MOCCA, the combination of statistical and N-body treatments allows for realistic results without compromising computational efficiency. Furthermore, simulations performed with codes based on a Monte Carlo approach, such as MOCCA, provide accurate results comparable to those obtained from full N-body simulations (see~\cite{rodriguez_morscher_2016} and the references therein), making them a powerful tool for exploring the evolution of globular clusters and their internal dynamics.

The clusters simulated for the MOCCA Survey 1 start with a Kroupa stellar initial mass function~\citep{kroupa_2001}, with a maximum stellar mass of $100\,{\rm M_{\odot}}$ and a minimum of $0.08\,{\rm M_{\odot}}$. The different clusters also have different metallicities and initial binary fractions; the details of the parameters used to simulate these clusters can be found in \cite{askar_szkudlarek_2016}. From Survey 1, we randomly choose one hundred BH binary systems in which at least one component is an IMBH with an initial mass $>500\,{\rm M_{\odot}}$ and with semi-major axes $p \lesssim 1\times10^4\,r_s$, where $r_s := 2Gm_{\rm IMBH}/c^2$ is the Schwarzschild radius of the IMBH -- with $m_{\rm IMBH}$ the mass of the IMBH and $c$ the speed of light. Note that for our analysis, we do not intend to consider a specific GC but a distribution of realistic orbital parameters of IMRIs and their host systems with different formation and interaction histories; therefore, we use this data for the initial configuration of the IMRIs we will analyze as well as for the properties of the host GC.

\subsection{Time between two encounters}\label{sec:encounters}

To model the Brownian motion, it is necessary to determine how often the IMRI encounters a star. We determine the free length of pass $s_F$ by integrating $N\eta(r)$ over the angular coordinates and in a radius from $r$ to $r+s_F$, where $N$ is the number of stars in the cluster. We set this integral to be equal to one, which means the IMRI encounters one star after pacing the length $s_F$. If we further assume that $s_F\ll r$, we find
\begin{equation}\label{eq:free}
    s_F(r_\ast) = \frac{16}{27}\frac{R_h}{N}\frac{(r_\ast^2 + 9/16)^{5/2}}{r_\ast^2},
\end{equation}
where we used \eq{eq:size} to replace $b$ by $R_h$ and we define $r_\ast:=r/R_h$. $s_F(r)$ diverges as $r$ approaches zero, but due to the displacement induced by the Brownian motion, the IMRI is expected never to be exactly at the center of the cluster. Moreover, as long as $r_\ast\gg\sqrt{R_h/(128N)}$, the assumption $s_F\ll r$ is fulfilled and thus we can use \eq{eq:free} for all radii considered in this study.

The time between two encounters is given by the free path length divided by the velocity with which the IMRI moves. We can approximate the velocity of the IMRI at a radius $r_\ast$ by using the virial theorem and considering the mass enclosed in that radius to get~\citep{binney_tremaine_2008}
\begin{equation}\label{eq:vel}
    v(r_\ast) = \sqrt{\frac{GM}{R_h}}\frac{r_\ast}{(r_\ast^2 + 9/16)^{3/4}},
\end{equation}
where $G$ is the gravitational constant and the mass enclosed in $r_\ast$ was obtained by integrating $M\eta(r_\ast)$ in spherical coordinates from zero to $r_\ast$. Combining \eq{eq:free} and \eq{eq:vel}, we then find for the time between two encounters
\begin{equation}\label{eq:encounter}
    t_{\rm enc}(r_\ast) = \frac{16}{27}\frac{R_h^{3/2}}{\sqrt{GMN^2}}\frac{(r_\ast^2 + 9/16)^{13/4}}{r_\ast^3}.
\end{equation}

\begin{figure}[tpb] \centering \includegraphics[width=0.48\textwidth]{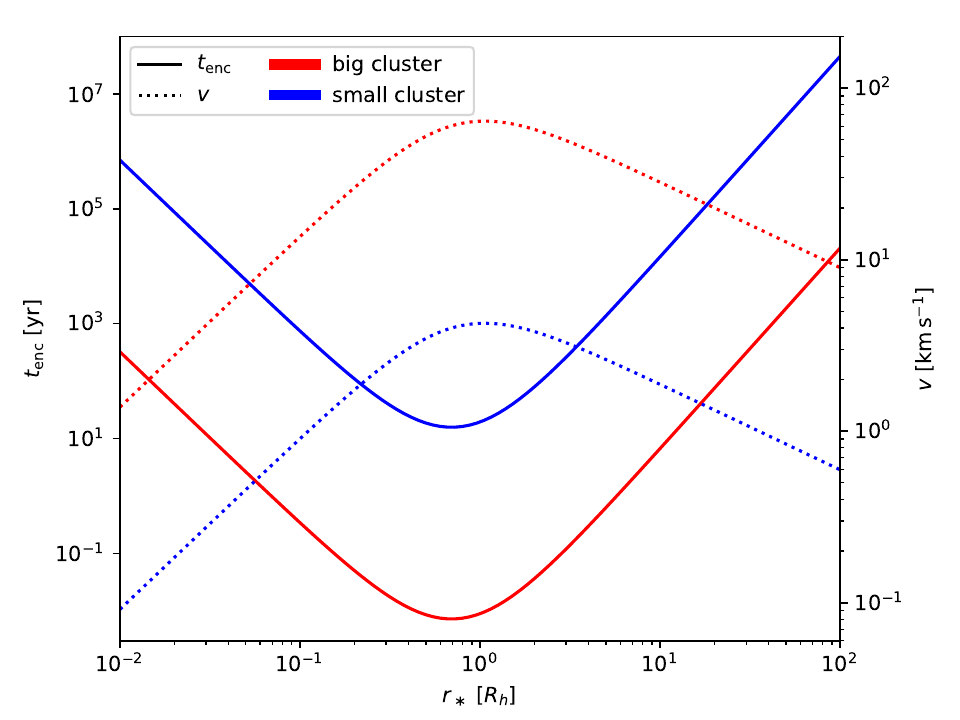}
\caption{
    The time between two encounters $t_{\rm en}$ (solid line) and the velocity of the IMRI (dotted line) as functions of the distance from the center of the star cluster $r_\ast$. We show these functions for a `big cluster' (red) and a 'small cluster' (blue). See the text for more details.
    }
\label{fig:timevel}
\end{figure}

In \fref{fig:timevel}, we show the time between two encounters $t_{\rm enc}$ as a function of $r$ for two cases: (i) a `small cluster' with $M=3.18\times10^4\,{\rm M_\odot}$, $R_h=3.81\,{\rm pc}$, and $N=8.14\times10^4$ and (iii) a `big cluster' with $M=1.23\times10^6\,{\rm M_\odot}$, $R_h=0.66\,{\rm pc}$, and $N=2.04\times10^6$ (cf. \tref{tab:src}). We see that $t_{\rm enc}$ has a strong dependence on $r$. For a big cluster, $t_{\rm enc}$ can be as small as $~10^{-2}\,{\rm yr}$ for $r\approx1\,R_h$ but increase up to several $100\,{\rm yr}$ for smaller radii of the order $10^{-2}\,R_h$. For bigger radii, $t_{\rm enc}$ increases even more, reaching more than ten thousand years for $r\approx10^{2}\,R_h$. For a small cluster, we see a similar behavior but with a shift of the minimum $t_{\rm enc}$ to longer times of over ten years at a radius of around $1\,R_h$. In the inner regions of the small cluster, $t_{\rm enc}$ goes up to several thousand to hundred thousand years, while in the outer regions, it goes up to some $10^7\,{\rm yr}$.

Note that $t_{\rm enc}$ is derived from $s_F$, which is calculated using the probability of the IMRI encountering one star when moving radially without specifying how close the encounter has to be. Only accounting for encounters due to a change in the radial direction could lead to an underestimation of the number of encounters; however, also counting far-away encounters boosts the number of possible encounters. We see from \fref{fig:timevel} that the number of encounters can be high in some cases, and thus it is unlikely that we are (significantly) underestimating their number -- at least in the most relevant regions where the number of encounters is high. In our study, we limit the interaction of the IMRI with the encountered star to be weak to guarantee a negligible perturbation of the inner properties of the IMRI. In \sref{sec:acceleration}, we discuss the encounters considered in more detail and estimate their duration independent from $t_{\rm enc}$ and $s_F$. As we will show, $t_{\rm enc}$ is always significantly bigger than the duration of a single encounter, showing that our estimation is not (significantly) overestimating the number of possible encounters.

The Brownian motion is a random process; hence, it is impossible to predict the radii at which the IMRI is at a given time. However, the displacement from the cluster's center $\Delta r$ can be estimated to be proportional to the square root of the velocity of the IMRI times $t_{\rm enc}$: $\Delta r\sim\sqrt{vt_{\rm enc}}$. We see from \fref{fig:timevel} that the trend of the velocity is inverse to the trend of $t_{\rm enc}$, having the highest velocities around $1\,R_h$. However, $t_{\rm enc}$ decreases faster than the velocity increases, and their product has a minimum around $1\,R_h$. Therefore, the displacement from the cluster's center slows significantly down when approaching $1\,R_h$, and it is reasonable to assume the IMRI will stay inside this `boundary'. Therefore, we conclude that for a small cluster, we expect an IMRI to have a close encounter $\mathcal{O}(10\text{-}10^5)\,{\rm yr}$ while for a big cluster, we expect a close encounter $\mathcal{O}(10^{-2}\text{-}10^2)\,{\rm yr}$.

\subsection{Acceleration of the IMRI}\label{sec:acceleration}

The effect of the Brownian motion observable using GWs is the acceleration induced by the interaction of the IMRI with a star during a close encounter. The acceleration the IMRI experiences when encountering a star is given by
\begin{equation}\label{eq:acc}
    a = \frac{Gm_\ast}{\alpha^2},
\end{equation}
where $m_\ast$ is the mass of the star encountered, and $\alpha$ is the distance between the IMRI and the star during the encounter.

The encounter with the star can change the parameters of the IMRI when it is very close, inducing multiple effects that range from altering the waveform to completely disrupting the binary~(see, e.g., \cite{antonini_rodriguez_2018,wen_2003,mandel_brown_2008}). We only want to consider encounters where the IMRI is not disrupted, and want to focus on the center of mass acceleration as the main observable feature. Therefore, we compute a minimal distance for the encounter $\alpha_{\rm min}$, by setting that the change in the specific kinetic energy of the IMRI due to the encounter is smaller than 1\,\% its specific binding energy $\Delta\mathcal{E}_{\rm kin}\lesssim0.01\mathcal{E}_{\rm bind}$.

The change in specific kinetic energy is given by~\citep{binney_tremaine_2008}
\begin{equation}\label{eq:kinen}
    \Delta\mathcal{E}_{\rm kin} = \frac12\left(\frac{2GM}{N\alpha_{\rm min}v}\right)^2,
\end{equation}
where $v$ is the velocity of the IMRI given in \eq{eq:vel} and we used the approximation that an average star has a mass $m_\ast = M/N$. For the specific binding energy, we use
\begin{equation}\label{eq:binen}
    \mathcal{E}_{\rm bin} = \frac{Gm_{\rm IMBH}}{2p},
\end{equation}
where $p$ is the semi-major axis of the binary at the time of the encounter, and we approximate the total mass of the system by the mass of the IMBH $m_{\rm IMBH}$. We can write $p=dr_s$, where $r_s$ is the Schwarzschild radius of the IMBH and $d$ is some proportionality factor. Combining these equations, we get
\begin{equation}\label{eq:almin}
    \alpha_{\rm min} \gtrsim \frac{\sqrt{800GMR_hd}}{Nc}\frac{(r_\ast^2 + 9/16)^{3/4}}{r_\ast}.
\end{equation}

Using \eq{eq:almin} and $m_\ast = M/N$ in \eq{eq:acc}, we find the upper limit for the acceleration of an IMRI interacting with a star but that is not broken by this interaction
\begin{equation}\label{eq:acceleration}
    a(r_\ast) \lesssim \frac{Nc^2}{800R_hd}\frac{r_\ast^2}{(r_\ast^2 + 9/16)^{3/2}}.
\end{equation}
The acceleration of the IMRI $a$ is limited by how closely the IMRI can get to the star encoded by the value of $d$. Other than that, the value of $a$ is dictated by the properties of the cluster $R_h$ and $N$ as well as its profile described by the second fraction in \eq{eq:acceleration}.

\begin{figure}[tpb] \centering \includegraphics[width=0.48\textwidth]{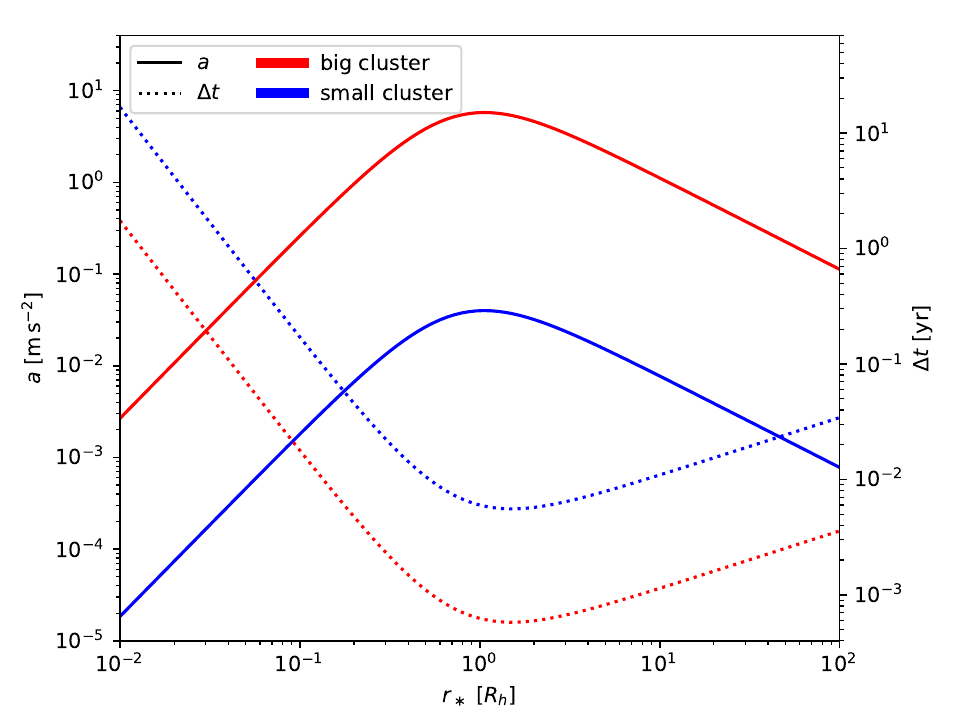}
\caption{
    The acceleration $a$ of an IMRI due to its Brownian motion (solid line) and the duration of an encounter $\Delta t$ (dotted line) as functions of the distance from the center of the star cluster $r_\ast$. $a$ and $\Delta t$ are shown for a `big cluster' (red) and a 'small cluster' (blue). See the text for more details.
    }
\label{fig:accdur}
\end{figure}

In \fref{fig:accdur}, we show the acceleration as a function of the radius at which the IMRI is for a small cluster and a big cluster, where we assume $d=1000$ (cf. \tref{tab:src}). We see that at a radius of around $1\,R_h$ the acceleration has a maximum of the order $1\,{\rm m\,s^{-2}}$ and $10^{-2}\,{\rm m\,s^{-2}}$ for a big cluster and a small cluster, respectively. For smaller radii, the acceleration decreases going down to $\sim10^{-1}\,{\rm m\,s^{-2}}$ and $\sim10^{-3}\,{\rm m\,s^{-2}}$ for a big cluster at a radius of $10^{-1}\,R_h$ and $10^{-2}\,R_h$, respectively, while for a small cluster and the same radii, the acceleration even goes down to $\sim10^{-3}\,{\rm m\,s^{-2}}$ and $\sim10^{-5}\,{\rm m\,s^{-2}}$, respectively.

The interaction of an IMRI with a field star leads to its acceleration as estimated in \eq{eq:acceleration} ultimately causing a change in its velocity. We consider close encounters but focus on the Brownian motion of the IMRI. Therefore, we are interested in encounters that lead to a change of the IMRI's velocity and, in particular, its direction, but that do not lead to the ejection of the IMRI from the cluster. Making the approximation that the velocity of the IMRI after the interaction is $v_{\rm int}\approx a(r_\ast)\Delta t$ and assuming that $v_{\rm int}$ is smaller than the escape velocity
\begin{equation}\label{eq:esc}
    v_{\rm esc}(r_\ast) = \sqrt{\frac{2GM}{R_h}}\left(r_\ast^2 + 9/16\right)^{-1/4},
\end{equation}
we can estimate the duration of the interaction $\Delta t$. Combining \eq{eq:acceleration} with \eq{eq:esc}, we get
\begin{equation}\label{eq:duration}
    \Delta t(r_\ast) \gtrsim \frac{800d}{Nc^2}\sqrt{2GMR_h}\frac{(r_\ast^2 + 9/16)^{5/4}}{r_\ast^2}.
\end{equation}
The duration of the encounter depends -- like the acceleration -- on $d$, which describes the state of the IMRI, and on the properties of the cluster as well its position inside the cluster. We show in \fref{fig:accdur} the estimated duration of an encounter as a function of the position of the IMRI for a big cluster and a small cluster, where we assume again $d=1000$. For a big cluster, $\Delta t$ ranges from around $1\,{\rm yr}$ for $r\approx10^{-2}\,R_h$ to $\sim10^{-3}\,{\rm yr}$ for $r\approx1\,R_h$. For a big cluster, the duration is approximately one order of magnitude bigger ranging from roughly $10\,{\rm yr}$ for a radius of $10^{-2}\,R_h$ down to an order of $10^{-2}\,{\rm yr}$ at around $1\,R_h$. In particular, we have that for both types of clusters and all radii, the duration of an encounter $\Delta t$ is much smaller than the time between two encounters $t_{\rm enc}$, confirming that our estimations are sensible.

In our subsequent analysis, we estimate the acceleration to be equal to the right-hand side of \eq{eq:acceleration}, i.e., equal to its maximum value, while we assume the duration of the encounter to be equal to the right-hand side of \eq{eq:duration} and thus equal to its minimal value. As we will explain in \sref{sec:dat} and \sref{sec:eff} in more detail, the detectability of the acceleration and, thus, the Brownian motion depends on the product of the acceleration and the duration. Therefore, taking the two extreme values gives us a representative result and guarantees we do not overestimate the observability of the acceleration. Nevertheless, exploring how different acceleration profiles affect the detection in detail is necessary to get a more complete understanding of the effect of the encounters on the signal. Moreover, for simplicity, we assume the acceleration to be aligned with the direction of the velocity and to alternate between parallel and anti-parallel for subsequent encounters. As we assume the velocity to be randomly oriented relative to the line of sight, this assumption does not lead to a significant over- or underestimation of the effect of the Brownian motion on the GW signal. However, considering a more realistic distribution of the orientation of the acceleration in future studies is necessary to further refine the modeling of the Brownian motion.

\section{Gravitational wave data analysis}\label{sec:dat}

Before we discuss the effect the Brownian motion has on the GWs emitted by an IMRI and how it can be detected using them, we introduce some concepts of GW data analysis that are relevant to our study. Assuming we have a detected signal $h(t)$ and a model of the signal $h_m(t)$ in the time-domain, their noise-weighted inner product is defined as~\citep{finn_1992,sathyaprakash_schutz_2009}
\begin{equation}\label{eq:ipro}
    \langle h|h_m\rangle = 4\Re\left[\int_0^\infty\dd f\frac{\tilde{h}(f)^\ast\tilde{h}_m(f)}{S_n(f)}\right],
\end{equation}
where $\tilde{h}(f)$ and $\tilde{h}_m(f)$ are the Fourier transform of $h(t)$ and $h_m(t)$, respectively, an asterisk indicates the complex conjugate, and $\Re$ means the real part of the function. Moreover, $S_n(f)$ is the frequency-dependent sensitivity curve of the detector considered.

The optimal signal-to-noise ratio (SNR) $\rho$ is obtained when the model of the signal is identical to the detected signal~\citep{finn_1992,sathyaprakash_schutz_2009}
\begin{equation}\label{eq:snr}
    \rho := \sqrt{\langle h|h\rangle}.
\end{equation}
However, for two different waveforms $h$ and $h_m$, the usual way to quantify their difference is to evaluate their overlap or match
\begin{equation}\label{eq:match}
    \mathcal{M} := \frac{\left\langle h | h_m \right\rangle}{\sqrt{\left\langle h | h \right\rangle\left\langle h_m | h_m \right\rangle}},
\end{equation}
so that $\mathcal{M}=1$ when there is a perfect match between the detected signal and the model of the signal. While the match shows the similarity between two waveforms, whether we can distinguish these two waveforms depends additionally on how strong the signal is. Following \cite{cutler_vallisneri_2007}, we obtain a rule-of-thumb criterion for the two waveforms to be distinguishable
\begin{equation}\label{eq:dist}
\rho > \sqrt{\frac{D}{2(1-\mathcal{M})}},
\end{equation}
where $D$ is the number of parameters to be measured in the analysis. Throughout the paper, we assume $D=15$ for the number of parameters of an IMRI\footnote{The parameters considered are: (i) the primary mass, (ii) the secondary mass, (iii) the initial semi-major axis, (iv) the initial eccentricity, (v) the luminosity distance of the source, (vi) and (vii) the sky localization of the source, (viii) the inclination of the source, (ix) the angle between the velocity and the line of sight, (x) the angle between the velocity and the angular momentum of the orbit, (xi) the magnitude of the acceleration, (xii) the initial velocity of the source, (xiii) the duration of the acceleration, (xiv) the ratio of encounters, and (xv) the spin of the primary mass which is assumed to be zero in all cases.} Focusing on the phase shift induced by the Brownian motion $\delta\Phi$ as the reason for the difference between the two waveforms, we can get an even simpler, although less accurate, criterion for the two waveforms to be distinguishable~\citep{lindblom_owen_2008,torres-orjuela_chen_2020}
\begin{equation}\label{eq:distshi}
    \delta\Phi \sim 1/\rho.
\end{equation}
Note that the underlying expression derived in \cite{lindblom_owen_2008} is obtained expanding to linear order in the phase shift $\delta\Phi$. Therefore, it is necessary to assume that $\delta\Phi$ is small to get a meaningful estimation.

In the case of multiple independent detectors, a combined SNR can be obtained using the linearity of the noise-weighted inner product~\citep{isoyama_nakano_2018}
\begin{equation}\label{eq:snrtot}
    \rho^2_{\rm tot} = \sum_i \rho^2_i,
\end{equation}
where $\rho_i$ is the SNR of the independent detectors. Using the linearity of the noise-weighted inner product and the definition of the match in \eq{eq:match}, we can derive the combined match between the detected signal $h$ and the model signal $h_m$
\begin{equation}\label{eq:mattot}
    \mathcal{M}_{\rm tot} = \sum_i \frac{\rho_i\tilde{\rho}_i}{\rho_{\rm tot}\tilde{\rho}_{\rm tot}}\mathcal{M}_i.
\end{equation}
Here $\rho_i$ is the SNR of the signal $h$ in the $i$'th detector, $\tilde{\rho}_i$ is the SNR of the model $h_m$ in the same detector, $\rho_{\rm tot}$ and $\tilde{\rho}_{\rm tot}$ are the combined SNRs of the signal and the model, respectively, and $\mathcal{M}_i$ is the match between $h$ and $h_m$ in detector $i$. The rule-of-thumb criteria in \eq{eq:dist} and \eq{eq:distshi} can then be applied in the same way as for a single detector.

\section{The effect of the Brownian motion on the gravitational wave}\label{sec:eff}

For an accelerated GW source, there are two types of phase shift: (i) a time-dependent Doppler shift induced by the change of the velocity along the line of sight and (ii) the aberrational phase shift induced by the change of the velocity components perpendicular to the line of sight~\citep{torres-orjuela_chen_2020}. There are also other effects the motion of the IMRI can have on the GWs emitted, like a change of the amplitude or of the different modes~\citep{torres-orjuela_chen_2022,torres-orjuela_chen_2021}, but here we focus on detecting the Brownian motion due to its effect on the phase of the wave. The total phase evolution of the $n$'th harmonic of the GW can be described as
\begin{equation}
    \Phi_{n}(t) = \Phi_{0,n} + \Phi_{{\rm GW}, n} + \Phi_{{\rm Dop},n}(t) + n\Phi_{\rm Abe}(t),
\end{equation}
where $\Phi_{0,n}$ is the initial phase at the start of the observation, $\Phi_{{\rm GW}, n} = \int\dd t'\omega_n(t')$ is the leading order of the GW phase with $\omega_n$ being the frequency of the $n$'th harmonic, $\Phi_{{\rm Dop},n}(t)$ is the Doppler shifted phase evolution of the $n$'th harmonic, and $\Phi_{\rm Abe}(t)$ is the aberrational phase shift. Assuming the velocity of the source $v$ is always much smaller than the speed of light, we can write
\begin{equation}\label{eq:dop}
    \Phi_{{\rm Dop},n}(t) = \int\dd t' \beta(t')\cos(\theta)\omega_n(t'),
\end{equation}
where $\beta:=v/c$, $\theta$ is the angle between the velocity and the line of sight. For the aberrational phase shift, we have
\begin{equation}\label{eq:abe}
    \Phi_{\rm Abe}(t) = \mu\beta(t)\sqrt{\sin^2(\nu) - \mu^2(\cos(\theta)-\cos(\delta)\cos(\nu))^2}
\end{equation}
where $\nu$ is the angle between the velocity and the angular momentum of the source's orbit, $\delta$ is the angle between the line of sight and the angular momentum of the source, $\mu=\sin^{-1}(\delta)$, and we further assumed the source is seen from an inclination $\delta\gtrsim\pi/36$ where $\delta=0$ corresponds to a face-on source. Note that we require $|\nu-\delta|\le\theta$ for the line of sight, the velocity, and the orbital angular momentum to have sensible orientations. We see from \eq{eq:dop} that $\Phi_{{\rm Dop},n}$ does not only depend on the evolution of the velocity but also on the evolution of the frequency of the $n$'th harmonic. $\Phi_{\rm Abe}$, in contrast, only depends on the evolution of the velocity, although its contribution to the total phase of the $n$'th harmonic is proportional to $n$.

From \eq{eq:dop} and \eq{eq:abe}, we can estimate the phase shift $\delta\Phi$ to be at least of the order $at_{\rm tot}/c$ when assuming the acceleration is constant and that $t_{\rm tot}$ is the total time the source is accelerated. In this case, it is fair to assume that the phase shift is small and we can use \eq{eq:distshi} to estimate the SNR required to observe the phase shift
\begin{equation}\label{eq:snrshi}
    \rho \sim \frac{c}{at_{\rm tot}}.
\end{equation}
This estimation ignores that the Doppler shift is also proportional to the time-dependent evolution of the frequency, which can further increase the total phase shift compared to a source with no phase shift. Therefore, the phase shift estimated here can be considered a `minimal' phase shift, and consequently, \eq{eq:snrshi} estimates the maximal SNR required to detect the phase shift induced by the Brownian motion.

A crucial question that arises for the observation of the Brownian motion from IMRIs is whether the source will emit GWs for an extended period that surpasses the duration of the interaction. We can estimate the time to coalescence for the IMRI using Peters' timescale~\citep{vazquez-aceves_lin_2022}
\begin{equation}\label{eq:ttm}
    t_c = \frac{10}{64}\frac{p^4c^5}{G^3m_{\rm SCO}m_{\rm IMBH}^2}g(e),
\end{equation}
where
\begin{equation}\label{eq:eccdep}
    g(e) = (1-e^2)^{7/2}\left[1+\frac{73}{24}e^2+ \frac{37}{96} e^4\right]^{-1}.
\end{equation}
We define $m_{\rm SCO}$ as the mass of the small compact object (SCO), $m_{\rm IMBH}$ as the mass of the IMBH, $p$ as the semi-major axis of the binary at the start of the observation, and $e$ as the initial eccentricity. The time the IMRI can be observed increases the more massive the IMBH is and the further away it is from the merger, while it decreases if the mass of the SCO increases and for increasing eccentricities. However, in his original work~\citep{peters_1964}, Peters assumes that the secular evolution of the eccentricity can be neglected, leading to inaccurate results for highly eccentric and relativistic systems. Therefore, the timescale in \eq{eq:ttm} must be corrected using a set of factors $R$, $Q_{\rm h}$ and $Q_{\rm s}$~\citep{zwick_capelo_2020,zwick_capelo_2021}. By multiplying \eq{eq:ttm} by the factor $R$, Peters' timescale is corrected for secular eccentricity evolution, and when multiplied by $Q_{h}$ and $Q_{s}$, 1.5 Post-Newtonian hereditary fluxes and spin-orbit couplings are added. The correction by $Q_{s}$ is, however, irrelevant for us as we ignore the effect of the spin.

Most of the IMRIs we consider are far from the merger, needing up to millions of years before the components collide (cf. \tref{tab:src}). Only a restricted number of less than 10\,\% will merge in the order of years or below. For the sources to be observed, they need to emit GWs in the band of the detector; however, most sources have very high eccentricities of more than $0.5$ and thus emit a significant amount of gravitational radiation at high frequencies far exceeding the orbital frequency of the binary. Therefore, the majority of sources are detectable in the ${\rm mHz}$-band of space-based LI detectors, and a significant number are also detectable by AI detectors in the ${\rm dHz}$-band long before they merge. Therefore, for most sources, we expect that the time they can be observed is mainly restricted by the time the detector is online.

The duration of an encounter varies strongly depending on the position of the IMRI in the cluster (cf. \fref{fig:accdur}), ranging from several years to only a few days. Therefore, depending on the properties of the cluster and the position of the IMRI in the cluster, we can expect to observe either (at most) one encounter with low acceleration that lasts for the entire observation time or multiple short encounters with high accelerations. We assume an observation time of $5\,{\rm yr}$~\citep{tianqin_2021,lisa_2024} and consider two cases: (i) the IMRI has one encounter with an acceleration of roughly $10^{-3}\,{\rm m\,s^{-2}}$ that lasts for the full five years, and (ii) the IMRI has a total of seven encounter each lasting for $10^{-2}\,{\rm yr}$ and having a magnitude of $10^{-1}\,{\rm m\,s^{-2}}$. Using that the minimal phase shift is proportional to the magnitude of the acceleration times the total duration of the encounter, we can use \eq{eq:snrshi} to estimate the maximal SNR required to detect the phase shifts induced. We find that for case (i), a maximal SNR of roughly 1900 is required for the detection of the phase shift, while for case (ii), a maximal SNR of around 1350 is needed. SNRs of more than 1000 are possible for IMRIs, and thus we conclude that the Brownian motion of IMRIs is detectable~\citep{tianqin_2024}. However, as we will see in \sref{sec:det}, where we study the detectability of the Brownian motion in more detail, significantly lower SNRs of a few tens can be enough to detect the phase shift in many realistic cases.

\section{Detection of the Brownian motion from gravitational wave phase shift}\label{sec:det}

In \sref{sec:brown} and \sref{sec:eff}, we discussed the encounters an IMRI will experience in a GC as well as the effect these encounters can have on the GWs emitted in a general manner. Now, we study the measurability of the Brownian motion using more realistic data obtained from MOCCA simulations. Details about the clusters and the IMRIs considered can be found in \tref{tab:rang} and \tref{tab:src} in the Appendix.

In \fref{fig:adt}, we show the acceleration during an encounter $a$, the ratio of encounters per year $\Gamma:={\rm yr}/t_{\rm enc}$, and the duration of an encounter $\Delta t$ an IMRI experiences, estimated for the clusters considered. Unsurprisingly, we see a strong correlation between high accelerations, short encounters, and frequent encounters, while low accelerations correlate with long infrequent encounters. The strong correlation between $a$ and $\Delta t$ is obvious as they are mutually dependent on each other. The correlation with $\Gamma$ is less obvious as it is derived independently from the two other quantities, but still plausible as it also strongly depends on the properties of the GC and the position of the IMRI in the cluster. We find that the acceleration varies between a few $10^{-6}\,{\rm m\,s^{-2}}$ and some $1\,{\rm m\,s^{-2}}$, while the ratio of encounters per year varies between less than $10^{-5}$ and up to some tens. The encounters last between a few hours and several hundred years.

\begin{figure}[tpb] \centering \includegraphics[width=0.48\textwidth]{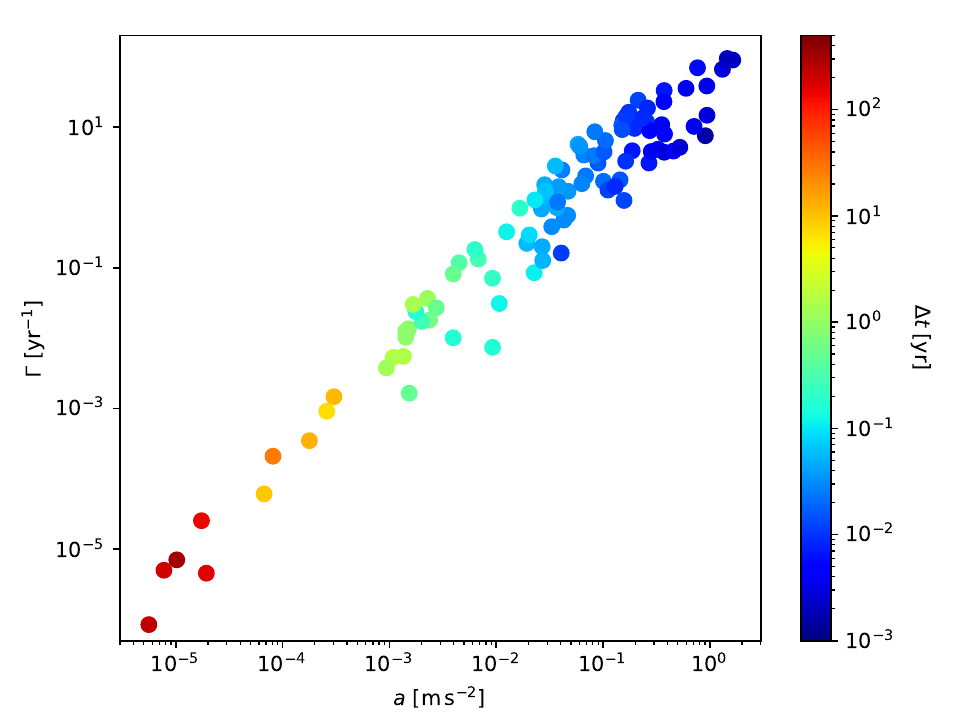}
\caption{
    The relevant parameters for the encounters of an IMRI in the different clusters considered. The abscissa shows the estimated acceleration during an encounter $a$, the ordinate, the estimated ratio of encounters per year $\Gamma$, and the color encodes the duration of an encounter $\Delta t$.
    }
\label{fig:adt}
\end{figure}

\subsection{Signal to noise ratio of intermediate mass ratio inspirals}\label{sec:snr}

We calculate the optimal SNR $\rho$ of all IMRIs using the equations introduced in \sref{sec:dat} for TianQin and LISA~\citep{tianqin_2021,lisa_2024}, as well as for AION~\citep{aion_2020}. We use the sensitivity curves of TianQin and LISA as described in \cite{torres-orjuela_huang_2023}, and for AION we use the sensitivity curve described in \cite{torres-orjuela_2024}. We include the displacement of TianQin and LISA due to their orbital motion, as well as the effect of the Earth's rotation on AION where we set its location to be close to London. For AION, we assume an on-off scheme of one year on and one year off, while for TianQin, we assume an on-off scheme of three months on and three months off, assuming an operation time of five years for both detectors. For LISA, we assume an operation time of three years with consecutive data collection.

The waveform model we use is similar to the one in \cite{vazquez-aceves_lin_2022} and \cite{vazquez-aceves_lin_2024}, which is based on \cite{barack_cutler_2004}. This waveform model was originally developed for sources with extreme mass ratios, however, recent studies such as \cite{wardell_pound_2023} have shown that waveform models developed for extreme mass ratios can achieve very good agreement with models developed for lower mass ratios. We modify the waveform model to include the effect of the time-dependent motion on the phase of the GW as well as on its amplitude~\citep{torres-orjuela_chen_2020,torres-orjuela_chen_2022}. Moreover, we only consider the $n_{0.1}$ modes with a power of at least 0.1\,\% of the power of the strongest mode to control the computation time. Nevertheless, for highly eccentric low-frequency sources, up to several tens of thousands of modes can be in the band of the detectors. In these cases, we limit the number of modes considered $n_{\rm lim}$ to be at most 1000 and only compute every $\Delta n$'th mode ($\Delta n := \lceil n_{0.1}/n_{\rm lim}\rceil$) in the waveform. When computing the noise-weighted inner product of a waveform with omitted modes, we compensate for them by multiplying the result by $\Delta n$; this approximation follows from the linearity of the noise-weighted inner product and the orthogonality of the modes combined with the fact that for waveforms with thousands of modes neighboring modes have similar power. We estimate the acceleration of the IMRI, as well as the frequency and duration of the encounters using the formulas derived in \sref{sec:brown}, where we use the initial semi-major axis of the IMRI for all encounters, ignoring its evolution in later encounters.

It is worth noting that the waveform model used only expands to leading-order in the eccentricity. In particular, when considering sources with initial eccentricities close to one, this approximation is not ideal. However, we checked that for all sources, the power emitted over all modes never diverges, guaranteeing the approximation to be at least mathematically sound. Moreover, the strong emission of GWs from highly eccentric sources results in these sources reducing their eccentricity relatively fast, and thus, the inaccuracies induced by a high eccentricity reduce quickly as the source/waveform evolves. In particular, we highlight that the leading-order expansion in the eccentricity can affect the SNR estimated, but not so much the match between a signal including the effect of the Brownian motion and a model waveform ignoring it, which is the main result of this paper. The reason is that the leading order approximation affects both scenarios the same way, and thus their difference is still led by the effect of the encounters have on the phase of the GWs.

\begin{figure}[tpb] \centering \includegraphics[width=0.48\textwidth]{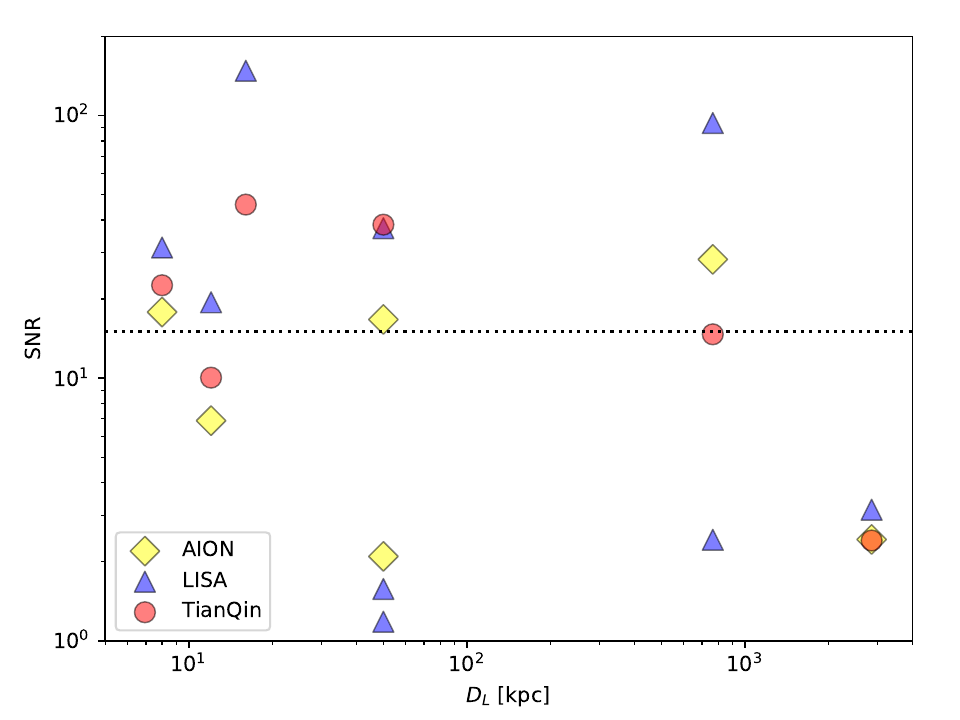}
\caption{
    The SNR of the IMRIs in AION (yellow diamonds), LISA (blue triangles), and TianQin (red circles) as a function of the source's luminosity distance $D_L$, where we only include sources with a SNR of at least 1. The black dotted line corresponds to the threshold SNR of 15.
    }
\label{fig:snr}
\end{figure}

\fref{fig:snr} shows the SNR of the IMRIs in AION, LISA, and TianQin as a function of the luminosity distance. Most sources have very low SNRs below one or are even outside the band of the detectors. We find that in AION only six sources have a SNR above one and just two surpass the estimated detection threshold of 15~\citep{arca-sedda_amaro-seoane_2021,torres-orjuela_huang_2023} -- the highest detected SNR being 28.3. In LISA, nine sources have an SNR of at least one, and five sources surpass the detection threshold, while in TianQin, only six sources have an SNR higher than one, and four of these sources have an SNR higher than 15. The highest SNR found in LISA is 148.0, while the highest SNR in TianQin is 45.8. See \tref{tab:det} in the appendix for more details. For all three detectors, the sources above the threshold are at most at the distance of the Andromeda Galaxy ($765\,{\rm kpc}$), and the highest SNR of a source in Maffei 1 ($2.85\,{\rm Mpc}$) is 3.2. However, we point out that this low detection distance is a result of most sources still being far from the merger. In \cite{torres-orjuela_huang_2023} and \cite{torres-orjuela_2024b}, it was shown that the detection distance of an IMRI can be significantly higher if it is close to the merger. We further point out that the closest source with an SNR above one is at a distance of $8\,{\rm kpc}$ despite considering five sources at a distance of only $4\,{\rm kpc}$. This result is, however, a consequence of the random distribution of the particular sources to the different locations. If any of the detectable sources were at a distance of $4\,{\rm kpc}$, we can expect that it would also have surpassed the threshold of 15.

\subsection{Match and detectability}\label{sec:match}

We have established that AION, LISA, and TianQin will detect a fraction of the IMRIs in GCs. The next question is whether these sources can be distinguished from IMRIs in the field or other environments. We consider the Brownian motion -- or in other words, multiple weak encounters inducing accelerations -- as the main observable. Therefore, we compare the waveform of an IMRI without encounters and one including encounters.

\begin{figure}[tpb] \centering \includegraphics[width=0.48\textwidth]{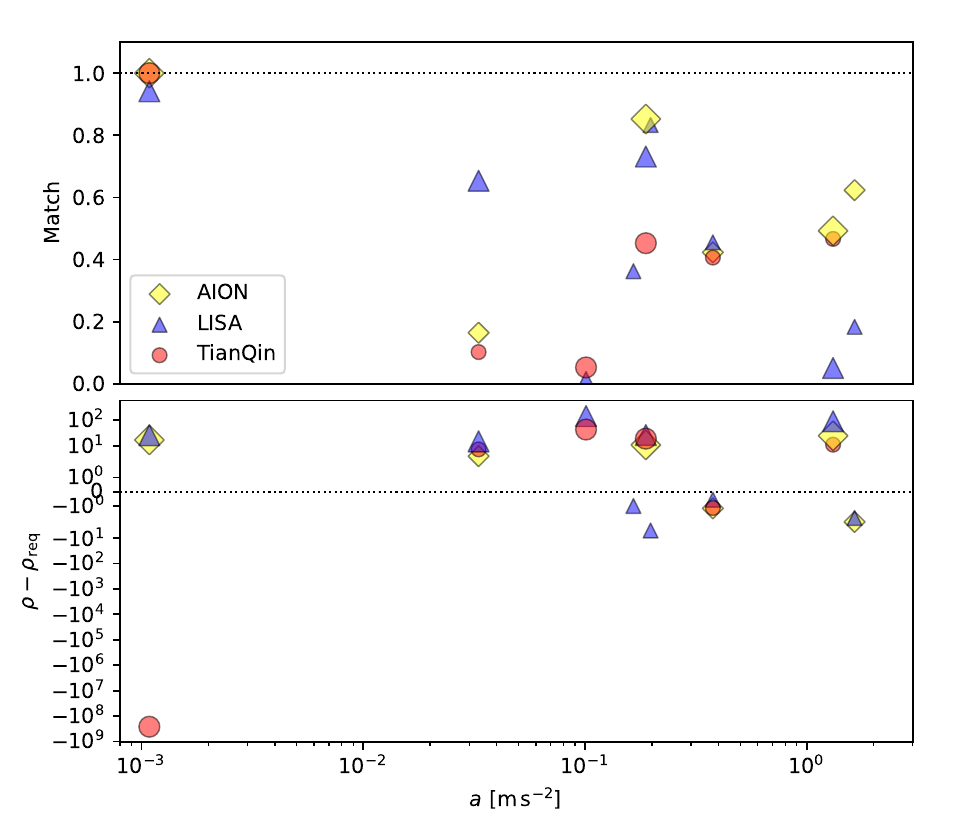}
\caption{
    The match (upper plot) and the difference between the SNR of the source and the SNR required for the detection of the acceleration (lower plot) as a function of the acceleration $a$. We show the values for AION (yellow diamonds), LISA (blue triangles), and TianQin (red circles), only including sources with a SNR of at least 1. We use bigger markers for sources that surpass the detection threshold.
    }
\label{fig:matcha}
\end{figure}

In \fref{fig:matcha}, we show the dependence on the magnitude of the acceleration. The upper subplot shows the match between the waveform with encounters and without for different accelerations. We see that only in a few cases will the match be close to one. For most sources and most detectors, the match goes even below 0.8, making a potential detection of the acceleration quite easy (cf. \tref{tab:det} for details). In the lower plot, we show the difference between the SNR of the source and the SNR required to detect the mismatch between the two waveforms, where the SNR required is estimated with \eq{eq:dist} assuming detection depends on $D=15$ different parameters.The majority of the sources have an SNR that is, in principle, high enough to distinguish between the two models, even when having an SNR of only one. Furthermore, for all sources that surpass the threshold SNR of 15, we are able to distinguish between the two waveform models, the only exception being one of the sources detected by TianQin. We point out that the lowest acceleration that can be detected and distinguished is of the order $10^{-3}\,{\rm m\,s^{_2}}$ and thus three orders of magnitude higher than the lowest acceleration we considered (cf. \fref{fig:adt}). However, whether this is a detection limit or just a coincidence because we have no source with a lower acceleration and a non-negligible SNR cannot be conclusively answered from our results.

\begin{figure}[tpb] \centering \includegraphics[width=0.48\textwidth]{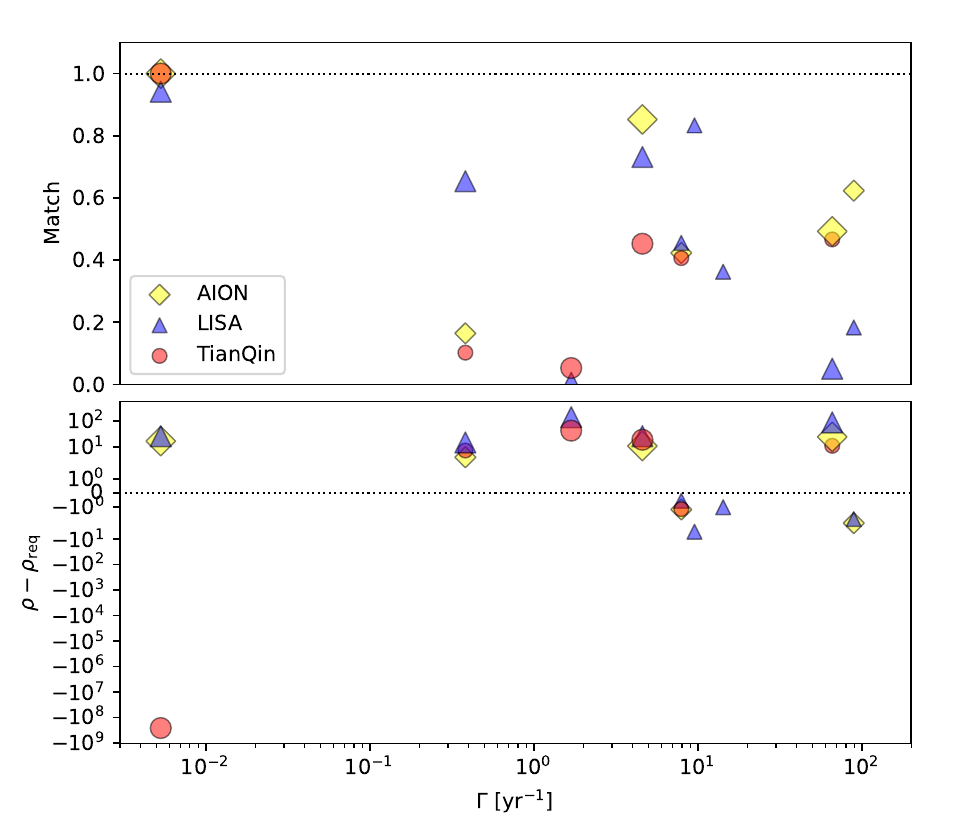}
\caption{
    The match (upper plot) and the difference between the SNR of the source and the SNR required for the detection of the encounters (lower plot) as a function of the ratio of encounters $\Gamma$. We use the same convention as in \fref{fig:matcha}.
    }
\label{fig:matche}
\end{figure}

We show the dependence of the detectability on the ratio of encounters in \fref{fig:matche}. Due to the strong correlation between the acceleration $a$ and the ratio of encounters $\Gamma$, \fref{fig:matche} looks similar to \fref{fig:matcha}. We find once again that most sources will have low matches below 0.8 with waveforms not including encounters. Moreover, we see that in most cases the SNR of the source is high enough to differentiate between the two waveforms -- in particular, if the source has an SNR higher than 15. We further find that the lowest encounter ratio detectable is of the order $10^{-3}\,{\rm yr^{-1}}$ and thus three orders of magnitude higher than the lowest encounter ratio considered. However, whether this is a real limit or just a coincidence due to our data can once again not be answered conclusively.

\subsection{Joint detection}\label{sec:joint}

It is possible that the operation time of AION, LISA, and TianQin will overlap (at least partially), allowing an improved analysis of the source due to joint detection. The benefits of joint detection include a higher SNR but also other advantages like better sky coverage, the reduction of data gaps, and different sensitivity at different frequencies (see \cite{torres-orjuela_huang_2023} for an extensive analysis in the case of TianQin-LISA joint detection). In this section, we discuss briefly the possible benefits of joint detection for four different cases: joint detection between AION and LISA, joint detection between AION and TianQin, joint detection between LISA and TianQin, and joint detection with all three detectors.

\begin{figure}[tpb] \centering \includegraphics[width=0.48\textwidth]{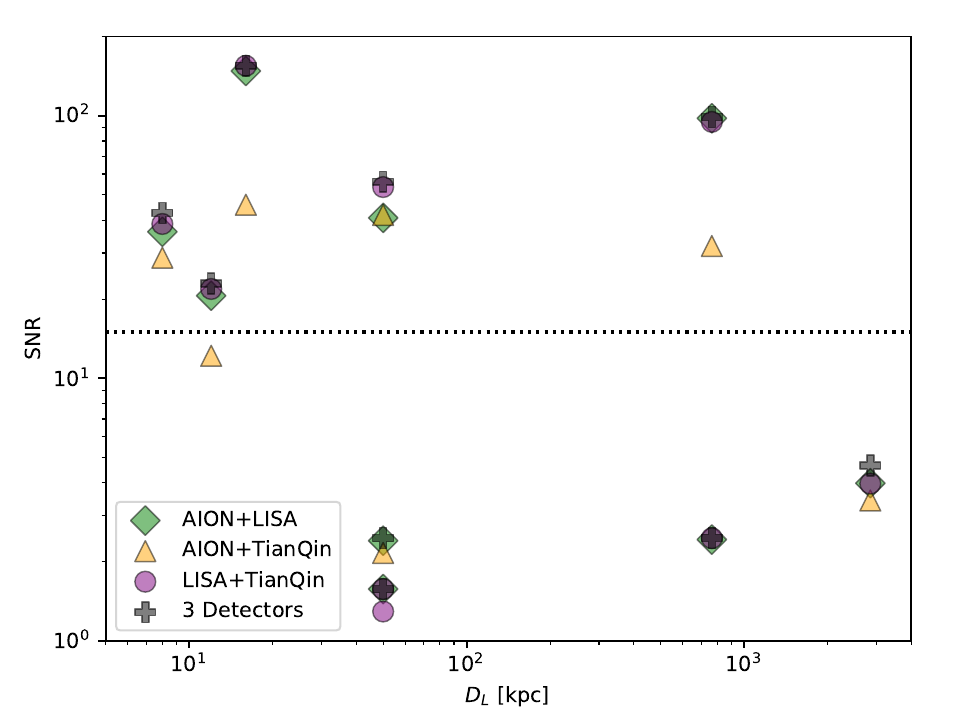}
\caption{
    The SNR of the IMRIs for joint detection between AION and LISA (green diamonds), AION and TianQin (orange triangles), LISA and TianQin (purple circles), and all three detectors (black crosses). We show the SNR depending on the luminosity distance of the source $D_L$, where we only include sources with a SNR of at least 1. The black dotted line corresponds to the threshold SNR of 15.
    }
\label{fig:snrj}
\end{figure}

In \fref{fig:snrj}, we show the SNR obtained from joint detection, where we used \eq{eq:snrtot} to calculate it. Obviously, the highest combined SNR is achieved when all three detectors are considered together. In most cases, the joint SNR from LISA and TianQin together is comparable to the joint SNR of all three detectors, while the joint detection between AION and LISA also gives comparable results. These results are plausible as all sources considered start relatively far from the merger and thus emit a significant amount of GWs at lower frequencies, where LISA is the most sensitive. However, in general, we can say that the difference between joint detection and single detection is not remarkable. Most sources that cannot be detected by a single detector will also not be detectable by joint detection. A more detailed analysis of joint detection, which goes beyond the scope of this work, is still interesting as a big fraction of the detectable sources show up in all or at least two detectors, and their different sensitivities allow for a more detailed analysis of different parameters of the source.

\begin{figure}[tpb] \centering \includegraphics[width=0.48\textwidth]{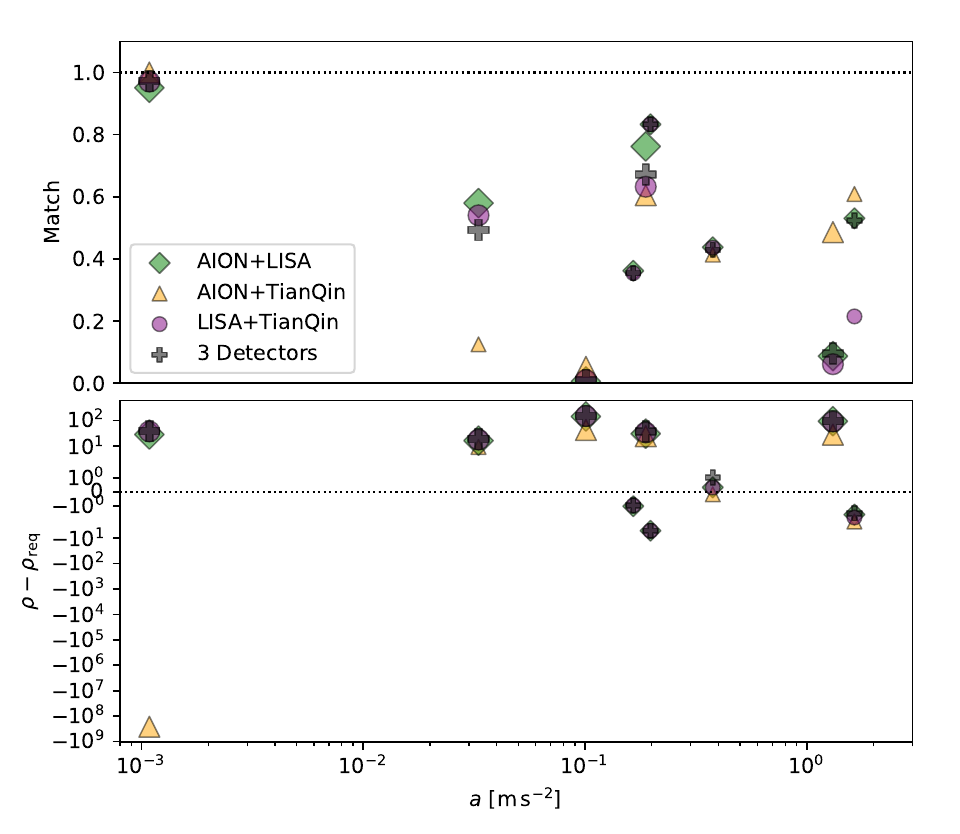}
\caption{
    The match (upper plot) and the difference between the SNR of the source and the SNR required for the detection of the acceleration (lower plot) as a function of the acceleration $a$. We show the values for joint detection between AION and LISA (green diamonds), AION and TianQin (orange triangles), LISA and TianQin (purple circles), and all three detectors (black crosses). We only include sources with an SNR of at least 1 and use bigger markers for sources that surpass the detection threshold.
    }
\label{fig:matchaj}
\end{figure}

\fref{fig:matchaj} shows the detectability of the acceleration from joint detection. Similar to the case of the SNR, joint detection allows for better detectability -- mainly due to the higher total SNR -- but the difference is not very prominent. Most remarkably, we find that in some cases the match between the model not containing the encounters and the one with encounters is sometimes higher for joint detection than for single detection. The reason is that in these cases the detector with the highest SNR also has the highest match, thus dominating the total match calculated from \eq{eq:mattot}. If, in contrast, the source with the highest SNR also has the lowest match, the total match goes down. We do not show a plot of the detectability of the ratio of encounters for joint detection, as the results obtained are very similar to those for the acceleration.

\subsection{Induced phase shift}\label{sec:phase}

\begin{figure}[tpb] \centering \includegraphics[width=0.48\textwidth]{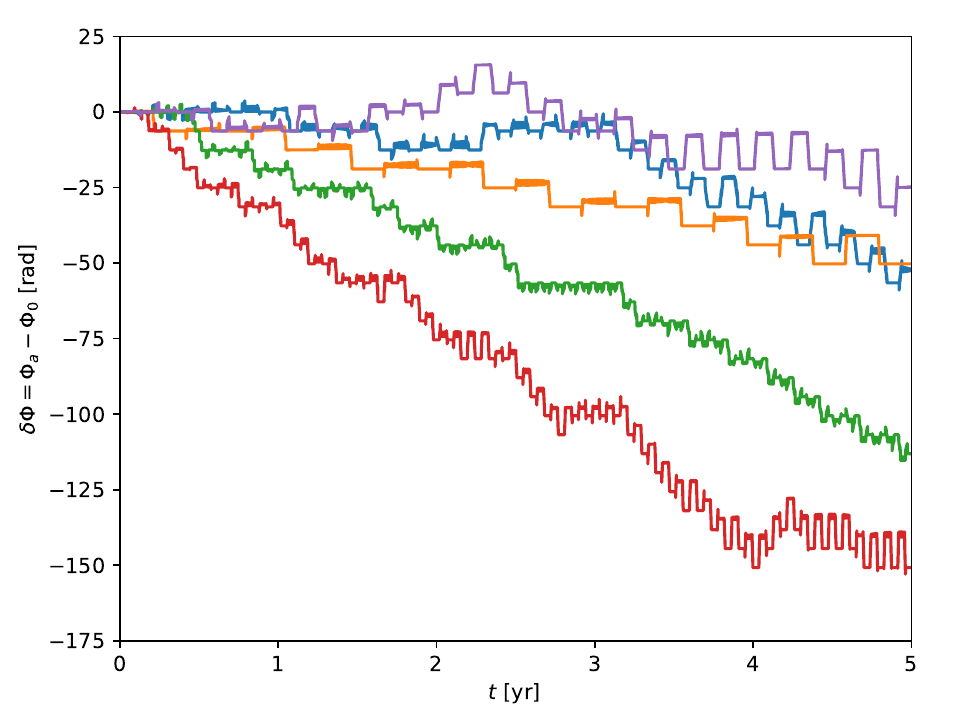}
\caption{
    The phase shift $\delta\Phi = \Phi_a - \Phi_0$ between the signal affected by the acceleration of the source ($\Phi_a$) and a waveform ignoring the effect of the encounters ($\Phi_0$) for some exemplary sources.
    }
\label{fig:phase}
\end{figure}

In the previous subsections, we showed that for several sources, the match between a signal affected by encounters and a model waveform ignoring the acceleration of the source can be very low. Due to this low match, low SNRs of just a few tens is enough to differentiate between the signal and the model; in fact, much lower than the theoretically estimated SNR at the end of \sref{sec:eff}. The main reason for this difference is that in the theoretical estimation, we considered the total phase shift to be proportional to the total change of velocity, ignoring the dependence of the Doppler shift on the evolution of the frequency. As we show in \fref{fig:phase}, the total phase shift $\delta\Phi := \Phi_a - \Phi_0$ between the phase of the signal $\Phi_a$ and the phase of the model waveform $\Phi_0$ can be over $150\,{\rm rad}$. This value is much greater than the theoretical estimation, which is always restricted to be smaller than one because the total change of the velocity must be smaller than the speed of light.

In \fref{fig:phase}, we further see that the phase shift does not evolve monotonically but can even change from positive to negative. The reason is that the direction of the acceleration can change the sign of the phase shift depending on its orientation relative to the line of sight. Moreover, the frequency's evolution is generally not linear; thus, the Doppler and aberrational phase shift can have different contributions to the total phase shift depending on the specific time they act on the signal emitted. Additional complexity is added to the phase shift by the random nature of the encounters. Our current model treats all encounters for a specific source to have the same magnitude, which is a strong simplification, but introduces encounters to the signal randomly. Therefore, we see for the examples in \fref{fig:phase} extended periods with rapidly increasing (decreasing) phase shifts alternating with periods where the phase shift only changes very little.

The complex behavior of the phase shift offers the possibility to differentiate the Brownian motion from other environmental effects inducing a phase shift, like the torques acting on GW sources embedded in gaseous environments. Gas drags tend to have a more uniform and non-randomly distributed effect on the phase of a GW source~\citep{garg_derdzinski_2022,speri_antonelli_2023}, and hence telling these two effects apart should be relatively easy. Only in the case where an IMRI experiences one weak encounter that lasts for the entire observation time, it might be more difficult to differentiate the Brownian motion from other, more regular effects. In this case, the gravitational interaction leads to a phase shift that is strongly time-dependent -- the time-dependent acceleration over long times results in highly non-linear Doppler and aberrational shifts -- and thus it should be possible to tell it apart from a more regular phase shift induced by gas. However, in our model, we treat the acceleration to be constant, and thus it is not possible to study this case in detail.

The random nature of encounters can also be used to study the properties of the host GC. As we show in \sref{sec:brown}, the parameters of the Brownian motion -- the magnitude of the acceleration, the duration of encounters, and the frequency of encounters -- depend on the properties of the cluster and the position of the IMRI in the cluster. Moreover, as can be seen in \fref{fig:adt}, there is a strong correlation between these properties, which should facilitate their measurement. At the same time, it is worth pointing out that the random nature of the Brownian motion makes it impossible to measure any property of the cluster with certainty. It is more realistic to assume that the detection of random encounters for multiple IMRIs will allow us to statistically prove the properties of GC. The encounters being randomly distributed and the strong correlation between the different parameters dictating the Brownian motion should also allow us to distinguish the effect of the Brownian motion from the evolution of the phase for an IMRI in a vacuum. However, studying this difference in more detail needs to be done in the future once we get accurate waveforms for IMRIs, and also requires further improving the model describing the Brownian motion.

\cite{tiwari_vijaykumar_2024a} study the acceleration of binary BHs in GCs, focusing on the acceleration induced by the gravitational potential of the GC. Although the accelerations induced by the potential are generally smaller than those induced by (weak) encounters, they find that it is still detectable in some cases. In \cite{tiwari_vijaykumar_2024b}, the same authors further explore how the detection of these accelerations can be used to study the environment of the host, finding that detailed information about the host environment can be directly inferred using GWs. Combining their results with a refined model of the Brownian motion should therefore allow for detailed studies of GCs and similar host systems.

\section{Conclusions}\label{sec:con}

In this paper, we study the detection of IMRIs in GCs by TianQin, LISA, and AION. In particular, we analyze the detectability of the Brownian motion of the IMRI induced by interactions with the stars in the cluster. The initial parameters of the considered IMRIs, as well as the properties of their host clusters, are taken from MOCCA simulations~\citep{askar_szkudlarek_2016}, thus representing realistic sources. We consider 100 different sources distributing their location to 25 systems inside the Milky Way, 25 systems at the location of the Large Magellanic Cloud, 25 systems in the Andromeda Galaxy, and 25 systems at the location of Maffei 1. For the Brownian motion, we focus on weak interactions with the surrounding stars that do not significantly alter the internal parameters of the star. We study the possibility of detecting these weak encounters by measuring the time-dependent phase shift of the GWs emitted induced by the acceleration of the IMRIs center of mass, where we consider both the Doppler shift and the aberrational phase shift~\citep{torres-orjuela_chen_2020}.

We calculate the SNR of the 100 sources in TianQin, LISA, and AION, finding that only a small fraction ($<10\,\%$) will have a SNR higher than 15, which is required for detection. A slightly bigger fraction, but still small ($\approx10\,\%$) will have a SNR of more than one, which means they could be detected if they were to be closer than the position assigned to them. However, the majority of the IMRIs will remain undetected due to their GWs being too weak and/or at too low frequencies. We further find that the biggest number of sources will be detected by LISA, closely followed by TianQin, as all sources are initially relatively far from the merger. AION is able to, in principle, see a similar number of sources, but a bigger fraction remains undetected due to their low SNR. The weak interactions with the surrounding stars (Brownian motion) lead to phase shifts that are significant in most cases and in some cases can even go up to more than $100\,{\rm rad}$ over the entire time of observation. A big fraction of the IMRIs have a low match ($\lesssim0.8$) with a waveform that does not include these interactions, and none of the sources with a SNR higher than one has a match equal to one in all detectors. In particular, all detectable sources have a mismatch that is big enough to be discerned with the actual SNR of the source, and even for those sources with an SNR between one and 15, the mismatch can be high enough for the interactions to become apparent. Moreover, we consider joint detection by pairs of the three detectors as well as by all three detectors together. The SNR obtained by joint detection is always higher, thus assisting the detection of the IMRIs and their Brownian motion. The improvement due to joint detection is not very prominent in the relatively simple analysis conducted in this work, but we suspect it to be more relevant in more detailed studies, including parameter estimation.

We find that a small fraction of IMRIs in GCs will be detectable by TianQin, LISA, and AION; although we plan to conduct a more detailed study with a bigger set of sources in future work. Furthermore, we show that the GW's phase shift induced by the Brownian motion of the IMRI due to weak interactions with surrounding stars is not only strong enough to be detectable but can significantly reduce the match between the actual signal and a model not including this effect. Therefore, we conclude that including the Brownian motion of IMRIs in future waveform models will be crucial for successful detection, but will also allow studying the properties of the GC hosting the IMRI.

\begin{acknowledgments}
ATO thanks Dieter Breitschwerdt and Stefan Harfst, who proposed the Brownian motion of massive black holes as the topic of his undergraduate thesis. Several ideas developed during that period were used in this paper. ATO further acknowledges support from the Key Laboratory of TianQin Project (Sun Yat-sen University), Ministry of Education (China). VVA acknowledges support from the Boya postdoctoral fellowship program of Peking University. We thank Mirek Giersz, Abbas Askar, and Arkadiusz Hypki for providing the MOCCA simulation data of the black hole binaries. 
\end{acknowledgments}

\appendix

\tref{tab:src} shows the properties of the sources considered including their ID, the mass of the IMBH $M_{\rm IMBH}$, the mass of the small compact object $m_{\rm SCO}$, the initial semi-major axis of the binary $p$ in Schwarzshild radii of the IMBH, the initial eccentricity of the binary $e$, the time to coalescence $t_c$ estimated with \eq{eq:ttm}, the total mass of the cluster $M$, the half-mass radius of the cluster $R_h$, the number of stars in the cluster $N$, and the location of the cluster. We consider eight different locations for the clusters: MW1 with a luminosity distance $D_L = 4\,{\rm kpc}$, a declination ${\rm dec} = +90^\circ$, and a right ascension ${\rm RA} = 7.8^{\rm h}$, MW2 with $D_L = 8\,{\rm kpc}$, ${\rm dec} = +45^\circ$, and ${\rm RA} = 7.2^{\rm h}$, MW3 with $D_L = 12\,{\rm kpc}$, ${\rm dec} = 36.7^\circ$, and ${\rm RA} = 0^{\rm h}$, MW4 with $D_L = 16\,{\rm kpc}$, ${\rm dec} = -90^\circ$, and ${\rm RA} = 10.8^{\rm h}$, MW5 with $D_L = 20\,{\rm kpc}$, ${\rm dec} = -45^\circ $, and ${\rm RA} = 11.5^{\rm h}$, LMC with $D_L = 50\,{\rm kpc}$, ${\rm dec} = -69.75^\circ$, and ${\rm RA} = 6.4^{\rm h}$, AG with $D_L = 765\,{\rm kpc}$, ${\rm dec} = +41.4^\circ$, and ${\rm RA} = 0.7^{\rm h}$, and M1 with $D_L = 2.85\,{\rm Mpc}$, ${\rm dec} = +59.7^\circ$, and ${\rm RA} = 2.6^{\rm h}$. MW1 to MW5 correspond to locations inside the Milky Way but are not related to an actual cluster. LMC corresponds to the location of the Large Magellanic Cloud, AG to the location of the Andromeda Galaxy, and M1 to the location of Maffei 1. Note that for the extragalactic host systems, we use the center of the galaxy as the location for the cluster, and for all locations, we use the center of the galaxy as the location of the source. As explained above, the IMRI does not reside at the center of the cluster due to its Brownian motion. Moreover, we consider GCs, which are usually not situated at the center of a galaxy. Nevertheless, the difference in location is so small that it does not alter our results, and thus we use the same locations to simplify the problem.

\begin{longtable}{| p{1cm} || p{1.8cm} | p{1.7cm} | p{1.5cm} | p{0.9cm} | p{1.7cm} || p{1.5cm} | p{1.3cm} | p{1.5cm} | p{1.2cm} |}
\caption{A list of the sources considered with their information: ID, mass of the IMBH $M_{\rm IMBH}$, mass of the small compact object $m_{\rm SCO}$, initial semi-major axis $p$, initial eccentricity $e$, time to coalescence $t_c$, total mass of the cluster $M$, half-mass radius of the cluster $R_h$, number of stars in the cluster $N$, and location of the cluster. See the text for details about the location.}\label{tab:src} \\
\hline
ID & $m_{\rm IMBH}\,{\rm [M_\odot]}$ & $m_{\rm SCO}\,{\rm [M_\odot]}$ & $p\,[r_s]$ & $e$ & $t_c\,{\rm [yr]}$ & $M\,{\rm [M_\odot]}$ & $R_h\,{\rm [pc]}$ & $N$ & location \\ \hline 
MWA & $ 4.32\times10^{3} $ & $ 4 $ & $ 2.95\times10^{3} $ & $ 0.817 $ & $ 2.76\times10^{6} $ & $ 3.02\times10^{5} $ & $ 2.89 $ & $ 8.8\times10^{5} $ & MW1 \\ \hline
MWB & $ 2.33\times10^{4} $ & $ 10 $ & $ 4.36\times10^{3} $ & $ 0.997 $ & $ 2.7\times10^{2} $ & $ 3.68\times10^{5} $ & $ 1.76 $ & $ 1.05\times10^{6} $ & MW1 \\ \hline
MWC & $ 2.65\times10^{4} $ & $ 13 $ & $ 7.92\times10^{3} $ & $ 0.964 $ & $ 9.79\times10^{6} $ & $ 3.05\times10^{5} $ & $ 2.59 $ & $ 9.1\times10^{5} $ & MW1 \\ \hline
MWD & $ 4.73\times10^{3} $ & $ 11 $ & $ 7.39\times10^{3} $ & $ 0.998 $ & $ 2.21\times10^{1} $ & $ 4.9\times10^{5} $ & $ 1.67 $ & $ 1.15\times10^{6} $ & MW1 \\ \hline
MWE & $ 5.63\times10^{2} $ & $ 46 $ & $ 4.32\times10^{3} $ & $ 0.672 $ & $ 1.13\times10^{5} $ & $ 4.78\times10^{5} $ & $ 3.52 $ & $ 1.09\times10^{6} $ & MW1 \\ \hline
MWF & $ 7.4\times10^{3} $ & $ 119 $ & $ 4.9\times10^{3} $ & $ 0.912 $ & $ 2.28\times10^{5} $ & $ 5.06\times10^{5} $ & $ 0.78 $ & $ 1.18\times10^{6} $ & MW2 \\ \hline
MWG & $ 1.27\times10^{4} $ & $ 11 $ & $ 2.16\times10^{3} $ & $ 0.635 $ & $ 1.92\times10^{7} $ & $ 4.11\times10^{5} $ & $ 2.53 $ & $ 1.11\times10^{6} $ & MW2 \\ \hline
MWH & $ 5.01\times10^{2} $ & $ 10 $ & $ 5.19\times10^{3} $ & $ 0.792 $ & $ 2.16\times10^{5} $ & $ 4.43\times10^{5} $ & $ 4.01 $ & $ 9.5\times10^{5} $ & MW2 \\ \hline
MWI & $ 1.18\times10^{3} $ & $ 29 $ & $ 2.62\times10^{3} $ & $ 0.968 $ & $ 7.03\times10^{1} $ & $ 5.22\times10^{5} $ & $ 2.93 $ & $ 1.08\times10^{6} $ & MW2 \\ \hline
MWJ & $ 1.5\times10^{3} $ & $ 16 $ & $ 4.95\times10^{3} $ & $ 0.926 $ & $ 4.13\times10^{4} $ & $ 4.89\times10^{5} $ & $ 0.77 $ & $ 1.18\times10^{6} $ & MW2 \\ \hline
MWK & $ 3.17\times10^{3} $ & $ 138 $ & $ 6.32\times10^{3} $ & $ 0.989 $ & $ 1.25\times10^{2} $ & $ 3.78\times10^{5} $ & $ 0.66 $ & $ 6.96\times10^{5} $ & MW3 \\ \hline
MWL & $ 8.72\times10^{2} $ & $ 12 $ & $ 2.3\times10^{3} $ & $ 0.965 $ & $ 7.41\times10^{1} $ & $ 3.27\times10^{5} $ & $ 3.41 $ & $ 6.93\times10^{5} $ & MW3 \\ \hline
MWM & $ 1.81\times10^{3} $ & $ 3 $ & $ 1.3\times10^{3} $ & $ 0.084 $ & $ 1.25\times10^{6} $ & $ 1.5\times10^{5} $ & $ 3.03 $ & $ 3.63\times10^{5} $ & MW3 \\ \hline
MWN & $ 8.19\times10^{2} $ & $ 19 $ & $ 3.05\times10^{3} $ & $ 0.318 $ & $ 8.87\times10^{5} $ & $ 3.2\times10^{5} $ & $ 1.08 $ & $ 7.27\times10^{5} $ & MW3 \\ \hline
MWO & $ 1.29\times10^{3} $ & $ 10 $ & $ 2.81\times10^{3} $ & $ 0.961 $ & $ 6.47\times10^{2} $ & $ 3.14\times10^{5} $ & $ 1.15 $ & $ 7.27\times10^{5} $ & MW3 \\ \hline
MWP & $ 2.25\times10^{3} $ & $ 10 $ & $ 4.96\times10^{3} $ & $ 0.991 $ & $ 1.2\times10^{2} $ & $ 3.09\times10^{5} $ & $ 1.21 $ & $ 7.25\times10^{5} $ & MW4 \\ \hline
MWQ & $ 9.71\times10^{2} $ & $ 4 $ & $ 1.78\times10^{3} $ & $ 0.047 $ & $ 9.31\times10^{5} $ & $ 1.87\times10^{5} $ & $ 4.78 $ & $ 5.2\times10^{5} $ & MW4 \\ \hline
MWR & $ 1.25\times10^{3} $ & $ 64 $ & $ 1.95\times10^{3} $ & $ 0.986 $ & $ 8.88\times10^{-1} $ & $ 5.5\times10^{5} $ & $ 1.31 $ & $ 1.11\times10^{6} $ & MW4 \\ \hline
MWS & $ 4.39\times10^{3} $ & $ 248 $ & $ 1.36\times10^{3} $ & $ 0.246 $ & $ 9.11\times10^{4} $ & $ 4.99\times10^{5} $ & $ 1.64 $ & $ 1.16\times10^{6} $ & MW4 \\ \hline
MWT & $ 1.33\times10^{4} $ & $ 31 $ & $ 1.7\times10^{3} $ & $ 0.609 $ & $ 3.72\times10^{6} $ & $ 3.95\times10^{5} $ & $ 3.02 $ & $ 1.12\times10^{6} $ & MW4 \\ \hline
MWU & $ 8.47\times10^{2} $ & $ 147 $ & $ 5.63\times10^{3} $ & $ 0.108 $ & $ 1.98\times10^{6} $ & $ 4.34\times10^{5} $ & $ 1.14 $ & $ 1.13\times10^{6} $ & MW5 \\ \hline
MWV & $ 1.58\times10^{3} $ & $ 80 $ & $ 4.8\times10^{3} $ & $ 0.489 $ & $ 2.53\times10^{6} $ & $ 6.03\times10^{4} $ & $ 2.26 $ & $ 1.44\times10^{5} $ & MW5 \\ \hline
MWW & $ 1.43\times10^{4} $ & $ 11 $ & $ 9.61\times10^{2} $ & $ 0.685 $ & $ 6.23\times10^{5} $ & $ 5.62\times10^{5} $ & $ 0.86 $ & $ 1.17\times10^{6} $ & MW5 \\ \hline
MWX & $ 7.42\times10^{2} $ & $ 23 $ & $ 5.08\times10^{3} $ & $ 0.181 $ & $ 6.03\times10^{6} $ & $ 5.91\times10^{5} $ & $ 3.06 $ & $ 1.2\times10^{6} $ & MW5 \\ \hline
MWY & $ 5.58\times10^{2} $ & $ 23 $ & $ 6.15\times10^{3} $ & $ 0.786 $ & $ 2.6\times10^{5} $ & $ 5.77\times10^{5} $ & $ 0.81 $ & $ 1.19\times10^{6} $ & MW5 \\ \hline
MCA & $ 6.23\times10^{2} $ & $ 25 $ & $ 4.23\times10^{3} $ & $ 0.684 $ & $ 2.03\times10^{5} $ & $ 5.63\times10^{5} $ & $ 0.85 $ & $ 1.19\times10^{6} $ & LMC \\ \hline
MCB & $ 1.01\times10^{3} $ & $ 32 $ & $ 7.85\times10^{3} $ & $ 0.969 $ & $ 3.33\times10^{3} $ & $ 5.54\times10^{5} $ & $ 0.89 $ & $ 1.19\times10^{6} $ & LMC \\ \hline
MCC & $ 2.84\times10^{4} $ & $ 23 $ & $ 7.8\times10^{3} $ & $ 0.959 $ & $ 9.52\times10^{6} $ & $ 4.02\times10^{5} $ & $ 1.91 $ & $ 1.02\times10^{6} $ & LMC \\ \hline
MCD & $ 1.04\times10^{4} $ & $ 12 $ & $ 7.99\times10^{2} $ & $ 0.533 $ & $ 4.46\times10^{5} $ & $ 4.89\times10^{5} $ & $ 2.19 $ & $ 1.14\times10^{6} $ & LMC \\ \hline
MCE & $ 5.24\times10^{2} $ & $ 26 $ & $ 5.67\times10^{3} $ & $ 0.824 $ & $ 8.2\times10^{4} $ & $ 3.15\times10^{5} $ & $ 0.96 $ & $ 6.81\times10^{5} $ & LMC \\ \hline
MCF & $ 6.51\times10^{2} $ & $ 23 $ & $ 3.84\times10^{3} $ & $ 0.996 $ & $ 1.63\times10^{-1} $ & $ 3.13\times10^{5} $ & $ 0.98 $ & $ 6.8\times10^{5} $ & LMC \\ \hline
MCG & $ 7.37\times10^{2} $ & $ 25 $ & $ 1.16\times10^{3} $ & $ 0.704 $ & $ 1.39\times10^{3} $ & $ 3.1\times10^{5} $ & $ 1.01 $ & $ 6.8\times10^{5} $ & LMC \\ \hline
MCH & $ 9.87\times10^{2} $ & $ 21 $ & $ 3.9\times10^{3} $ & $ 0.99 $ & $ 7.95 $ & $ 3.03\times10^{5} $ & $ 1.08 $ & $ 6.77\times10^{5} $ & LMC \\ \hline
MCI & $ 4.67\times10^{3} $ & $ 14 $ & $ 1.86\times10^{3} $ & $ 0.333 $ & $ 4.92\times10^{6} $ & $ 2.6\times10^{5} $ & $ 1.72 $ & $ 6.51\times10^{5} $ & LMC \\ \hline
MCJ & $ 5.65\times10^{2} $ & $ 12 $ & $ 7.51\times10^{3} $ & $ 0.4 $ & $ 1.81\times10^{7} $ & $ 3.18\times10^{4} $ & $ 3.81 $ & $ 8.14\times10^{4} $ & LMC \\ \hline
MCK & $ 2.92\times10^{4} $ & $ 26 $ & $ 2.31\times10^{3} $ & $ 0.457 $ & $ 1.68\times10^{8} $ & $ 5.12\times10^{5} $ & $ 2.7 $ & $ 1.62\times10^{6} $ & LMC \\ \hline
MCL & $ 2.92\times10^{4} $ & $ 26 $ & $ 7.63\times10^{3} $ & $ 0.959 $ & $ 8.06\times10^{6} $ & $ 5.12\times10^{5} $ & $ 2.7 $ & $ 1.62\times10^{6} $ & LMC \\ \hline
MCM & $ 4.11\times10^{4} $ & $ 41 $ & $ 5.93\times10^{3} $ & $ 0.905 $ & $ 5.53\times10^{7} $ & $ 5.18\times10^{5} $ & $ 2.64 $ & $ 1.6\times10^{6} $ & LMC \\ \hline
MCN & $ 1.52\times10^{3} $ & $ 57 $ & $ 3.15\times10^{3} $ & $ 0.895 $ & $ 6.18\times10^{3} $ & $ 1.23\times10^{6} $ & $ 0.66 $ & $ 2.04\times10^{6} $ & LMC \\ \hline
MCO & $ 5.43\times10^{3} $ & $ 163 $ & $ 2.04\times10^{3} $ & $ 0.968 $ & $ 1.08\times10^{2} $ & $ 1.12\times10^{6} $ & $ 0.78 $ & $ 2.07\times10^{6} $ & LMC \\ \hline
MCP & $ 1.77\times10^{4} $ & $ 23 $ & $ 9.78\times10^{2} $ & $ 0.542 $ & $ 1.42\times10^{6} $ & $ 9.03\times10^{5} $ & $ 1.36 $ & $ 1.97\times10^{6} $ & LMC \\ \hline
MCQ & $ 2.07\times10^{4} $ & $ 14 $ & $ 8.88\times10^{2} $ & $ 0.595 $ & $ 1.54\times10^{6} $ & $ 8.35\times10^{5} $ & $ 1.65 $ & $ 1.93\times10^{6} $ & LMC \\ \hline
MCR & $ 2.15\times10^{4} $ & $ 26 $ & $ 4.66\times10^{3} $ & $ 0.952 $ & $ 1.01\times10^{6} $ & $ 8.17\times10^{5} $ & $ 1.74 $ & $ 1.91\times10^{6} $ & LMC \\ \hline
MCS & $ 2.37\times10^{4} $ & $ 23 $ & $ 7.36\times10^{3} $ & $ 0.994 $ & $ 8.19\times10^{3} $ & $ 7.44\times10^{5} $ & $ 2.14 $ & $ 1.84\times10^{6} $ & LMC \\ \hline
MCT & $ 2.69\times10^{4} $ & $ 4 $ & $ 9.95\times10^{2} $ & $ 0.403 $ & $ 3.62\times10^{7} $ & $ 5.85\times10^{5} $ & $ 2.95 $ & $ 1.52\times10^{6} $ & LMC \\ \hline
MCU & $ 8.5\times10^{3} $ & $ 11 $ & $ 6.08\times10^{3} $ & $ 0.989 $ & $ 7.81\times10^{3} $ & $ 8.52\times10^{5} $ & $ 0.81 $ & $ 2.06\times10^{6} $ & LMC \\ \hline
MCV & $ 9.13\times10^{3} $ & $ 15 $ & $ 1.94\times10^{3} $ & $ 0.959 $ & $ 5.96\times10^{3} $ & $ 8.52\times10^{5} $ & $ 0.81 $ & $ 2.06\times10^{6} $ & LMC \\ \hline
MCW & $ 2.15\times10^{4} $ & $ 61 $ & $ 1.26\times10^{3} $ & $ 0.281 $ & $ 6.25\times10^{6} $ & $ 7.25\times10^{5} $ & $ 1.23 $ & $ 1.96\times10^{6} $ & LMC \\ \hline
MCX & $ 2.32\times10^{4} $ & $ 23 $ & $ 3.35\times10^{3} $ & $ 0.876 $ & $ 7.49\times10^{6} $ & $ 6.94\times10^{5} $ & $ 1.38 $ & $ 1.94\times10^{6} $ & LMC \\ \hline
MCY & $ 2.56\times10^{4} $ & $ 25 $ & $ 2.84\times10^{3} $ & $ 0.807 $ & $ 1.69\times10^{7} $ & $ 6.34\times10^{5} $ & $ 1.78 $ & $ 1.87\times10^{6} $ & LMC \\ \hline
AGA & $ 2.93\times10^{4} $ & $ 3 $ & $ 3.98\times10^{3} $ & $ 0.927 $ & $ 3.12\times10^{7} $ & $ 5.3\times10^{5} $ & $ 2.59 $ & $ 1.67\times10^{6} $ & AG \\ \hline
AGB & $ 3.82\times10^{3} $ & $ 275 $ & $ 1.98\times10^{3} $ & $ 0.991 $ & $ 4.5\times10^{-1} $ & $ 9.27\times10^{5} $ & $ 0.66 $ & $ 2.04\times10^{6} $ & AG \\ \hline
AGC & $ 1.51\times10^{4} $ & $ 36 $ & $ 5.18\times10^{3} $ & $ 0.972 $ & $ 9.6\times10^{4} $ & $ 8.28\times10^{5} $ & $ 0.86 $ & $ 2.04\times10^{6} $ & AG \\ \hline
AGD & $ 2.56\times10^{4} $ & $ 51 $ & $ 1.56\times10^{3} $ & $ 0.792 $ & $ 9.56\times10^{5} $ & $ 6.71\times10^{5} $ & $ 1.5 $ & $ 1.91\times10^{6} $ & AG \\ \hline
AGE & $ 3.12\times10^{4} $ & $ 13 $ & $ 1.45\times10^{3} $ & $ 0.599 $ & $ 2.63\times10^{7} $ & $ 5.16\times10^{5} $ & $ 2.68 $ & $ 1.62\times10^{6} $ & AG \\ \hline
AGF & $ 3.18\times10^{4} $ & $ 5 $ & $ 1.73\times10^{3} $ & $ 0.616 $ & $ 1.19\times10^{8} $ & $ 4.85\times10^{5} $ & $ 2.88 $ & $ 1.53\times10^{6} $ & AG \\ \hline
AGG & $ 1.75\times10^{4} $ & $ 12 $ & $ 2.06\times10^{3} $ & $ 0.873 $ & $ 1.19\times10^{6} $ & $ 8.05\times10^{5} $ & $ 0.92 $ & $ 2.03\times10^{6} $ & AG \\ \hline
AGH & $ 2.71\times10^{4} $ & $ 18 $ & $ 9.23\times10^{2} $ & $ 0.592 $ & $ 2.54\times10^{6} $ & $ 6.32\times10^{5} $ & $ 1.78 $ & $ 1.87\times10^{6} $ & AG \\ \hline
AGI & $ 3.17\times10^{4} $ & $ 27 $ & $ 5.11\times10^{3} $ & $ 0.999 $ & $ 3.5\times10^{-1} $ & $ 4.82\times10^{5} $ & $ 2.89 $ & $ 1.52\times10^{6} $ & AG \\ \hline
AGJ & $ 2.31\times10^{4} $ & $ 10 $ & $ 3.81\times10^{3} $ & $ 0.999 $ & $ 2.25 $ & $ 6.13\times10^{5} $ & $ 3.03 $ & $ 1.83\times10^{6} $ & AG \\ \hline
AGK & $ 1.56\times10^{3} $ & $ 12 $ & $ 1.33\times10^{3} $ & $ 0.234 $ & $ 2.16\times10^{5} $ & $ 6.02\times10^{5} $ & $ 3.75 $ & $ 1.5\times10^{6} $ & AG \\ \hline
AGL & $ 2.11\times10^{3} $ & $ 11 $ & $ 2.33\times10^{3} $ & $ 0.157 $ & $ 4.63\times10^{6} $ & $ 5.67\times10^{5} $ & $ 3.7 $ & $ 1.45\times10^{6} $ & AG \\ \hline
AGM & $ 6.56\times10^{3} $ & $ 24 $ & $ 3.87\times10^{3} $ & $ 0.996 $ & $ 1.02\times10^{1} $ & $ 4.36\times10^{5} $ & $ 3.59 $ & $ 1.21\times10^{6} $ & AG \\ \hline
AGN & $ 1.38\times10^{3} $ & $ 27 $ & $ 3.44\times10^{3} $ & $ 0.975 $ & $ 1.38\times10^{2} $ & $ 8.47\times10^{5} $ & $ 3.25 $ & $ 1.93\times10^{6} $ & AG \\ \hline
AGO & $ 7.44\times10^{2} $ & $ 26 $ & $ 3.44\times10^{3} $ & $ 0.724 $ & $ 8.57\times10^{4} $ & $ 8.68\times10^{5} $ & $ 0.73 $ & $ 2.05\times10^{6} $ & AG \\ \hline
AGP & $ 2.83\times10^{3} $ & $ 11 $ & $ 6.86\times10^{3} $ & $ 0.981 $ & $ 9.22\times10^{3} $ & $ 8.33\times10^{5} $ & $ 0.82 $ & $ 2.07\times10^{6} $ & AG \\ \hline
AGQ & $ 1.78\times10^{4} $ & $ 12 $ & $ 1.65\times10^{3} $ & $ 0.511 $ & $ 2.73\times10^{7} $ & $ 3.1\times10^{5} $ & $ 2.57 $ & $ 9.54\times10^{5} $ & AG \\ \hline
AGR & $ 2.45\times10^{4} $ & $ 25 $ & $ 7.88\times10^{3} $ & $ 0.969 $ & $ 2.52\times10^{6} $ & $ 3.26\times10^{5} $ & $ 2.32 $ & $ 9.69\times10^{5} $ & AG \\ \hline
AGS & $ 5.47\times10^{2} $ & $ 19 $ & $ 5.83\times10^{3} $ & $ 0.821 $ & $ 1.43\times10^{5} $ & $ 5.5\times10^{5} $ & $ 1.3 $ & $ 1.11\times10^{6} $ & AG \\ \hline
AGT & $ 1.28\times10^{4} $ & $ 11 $ & $ 5.62\times10^{3} $ & $ 0.949 $ & $ 2.14\times10^{6} $ & $ 4.19\times10^{5} $ & $ 2.53 $ & $ 1.14\times10^{6} $ & AG \\ \hline
AGU & $ 6.34\times10^{2} $ & $ 53 $ & $ 3.97\times10^{3} $ & $ 0.372 $ & $ 4.69\times10^{5} $ & $ 5.35\times10^{5} $ & $ 1.28 $ & $ 1.14\times10^{6} $ & AG \\ \hline
AGV & $ 7.94\times10^{2} $ & $ 25 $ & $ 4.45\times10^{3} $ & $ 0.921 $ & $ 6.12\times10^{3} $ & $ 5.15\times10^{5} $ & $ 1.42 $ & $ 1.13\times10^{6} $ & AG \\ \hline
AGW & $ 8.41\times10^{2} $ & $ 8 $ & $ 6.94\times10^{3} $ & $ 0.997 $ & $ 1.39 $ & $ 5.06\times10^{5} $ & $ 1.48 $ & $ 1.13\times10^{6} $ & AG \\ \hline
AGX & $ 9.76\times10^{2} $ & $ 8 $ & $ 3.54\times10^{3} $ & $ 0.381 $ & $ 4.19\times10^{6} $ & $ 4.77\times10^{5} $ & $ 1.69 $ & $ 1.12\times10^{6} $ & AG \\ \hline
AGY & $ 1.32\times10^{3} $ & $ 4 $ & $ 2.12\times10^{3} $ & $ 0.211 $ & $ 2.62\times10^{6} $ & $ 4.48\times10^{5} $ & $ 1.91 $ & $ 1.09\times10^{6} $ & AG \\ \hline
M1A & $ 1.52\times10^{3} $ & $ 22 $ & $ 4.99\times10^{3} $ & $ 0.483 $ & $ 1\times10^{7} $ & $ 4.24\times10^{5} $ & $ 2.09 $ & $ 1.05\times10^{6} $ & M1 \\ \hline
M1B & $ 8.4\times10^{2} $ & $ 23 $ & $ 1.61\times10^{3} $ & $ 0.958 $ & $ 1.7\times10^{1} $ & $ 5.34\times10^{5} $ & $ 1.29 $ & $ 1.15\times10^{6} $ & M1 \\ \hline
M1C & $ 8.64\times10^{2} $ & $ 89 $ & $ 7.99\times10^{3} $ & $ 0.618 $ & $ 2.47\times10^{6} $ & $ 5.28\times10^{5} $ & $ 1.33 $ & $ 1.14\times10^{6} $ & M1 \\ \hline
M1D & $ 1.12\times10^{3} $ & $ 28 $ & $ 5.7\times10^{3} $ & $ 0.607 $ & $ 3.68\times10^{6} $ & $ 4.98\times10^{5} $ & $ 1.56 $ & $ 1.13\times10^{6} $ & M1 \\ \hline
M1E & $ 1.98\times10^{3} $ & $ 10 $ & $ 1.8\times10^{3} $ & $ 0.187 $ & $ 1.53\times10^{6} $ & $ 3.77\times10^{5} $ & $ 2.42 $ & $ 9.54\times10^{5} $ & M1 \\ \hline
M1F & $ 9.54\times10^{2} $ & $ 8 $ & $ 5.12\times10^{3} $ & $ 0.845 $ & $ 3.85\times10^{5} $ & $ 4.42\times10^{5} $ & $ 1.06 $ & $ 1.16\times10^{6} $ & M1 \\ \hline
M1G & $ 2.66\times10^{3} $ & $ 28 $ & $ 4.4\times10^{3} $ & $ 0.464 $ & $ 1.64\times10^{7} $ & $ 3.87\times10^{5} $ & $ 1.56 $ & $ 1.12\times10^{6} $ & M1 \\ \hline
M1H & $ 2.04\times10^{3} $ & $ 25 $ & $ 5.14\times10^{3} $ & $ 0.822 $ & $ 9.06\times10^{5} $ & $ 4.49\times10^{5} $ & $ 1.02 $ & $ 1.17\times10^{6} $ & M1 \\ \hline
M1I & $ 2.65\times10^{3} $ & $ 10 $ & $ 3.31\times10^{3} $ & $ 0.806 $ & $ 8.37\times10^{5} $ & $ 4.06\times10^{5} $ & $ 1.39 $ & $ 1.14\times10^{6} $ & M1 \\ \hline
M1J & $ 2.68\times10^{3} $ & $ 20 $ & $ 3.31\times10^{3} $ & $ 0.8 $ & $ 4.74\times10^{5} $ & $ 4.04\times10^{5} $ & $ 1.4 $ & $ 1.14\times10^{6} $ & M1 \\ \hline
M1K & $ 2.86\times10^{3} $ & $ 25 $ & $ 5.49\times10^{3} $ & $ 0.824 $ & $ 2.19\times10^{6} $ & $ 3.96\times10^{5} $ & $ 1.48 $ & $ 1.13\times10^{6} $ & M1 \\ \hline
M1L & $ 3.56\times10^{3} $ & $ 7 $ & $ 6.25\times10^{2} $ & $ 0.543 $ & $ 3.19\times10^{4} $ & $ 3.76\times10^{5} $ & $ 1.7 $ & $ 1.1\times10^{6} $ & M1 \\ \hline
M1M & $ 5.58\times10^{2} $ & $ 19 $ & $ 5.62\times10^{3} $ & $ 0.715 $ & $ 5.05\times10^{5} $ & $ 5.65\times10^{5} $ & $ 1.09 $ & $ 1.15\times10^{6} $ & M1 \\ \hline
M1N & $ 5.94\times10^{2} $ & $ 24 $ & $ 6.69\times10^{3} $ & $ 0.853 $ & $ 1.28\times10^{5} $ & $ 5.65\times10^{5} $ & $ 1.09 $ & $ 1.15\times10^{6} $ & M1 \\ \hline
M1O & $ 6.63\times10^{2} $ & $ 21 $ & $ 7.83\times10^{3} $ & $ 0.993 $ & $ 1.92\times10^{1} $ & $ 5.59\times10^{5} $ & $ 1.12 $ & $ 1.15\times10^{6} $ & M1 \\ \hline
M1P & $ 8.98\times10^{2} $ & $ 19 $ & $ 3.79\times10^{3} $ & $ 0.885 $ & $ 1.77\times10^{4} $ & $ 5.42\times10^{5} $ & $ 1.23 $ & $ 1.14\times10^{6} $ & M1 \\ \hline
M1Q & $ 1\times10^{3} $ & $ 14 $ & $ 4.48\times10^{3} $ & $ 0.799 $ & $ 3.19\times10^{5} $ & $ 5.36\times10^{5} $ & $ 1.27 $ & $ 1.14\times10^{6} $ & M1 \\ \hline
M1R & $ 1.02\times10^{3} $ & $ 24 $ & $ 7.38\times10^{3} $ & $ 0.854 $ & $ 5.39\times10^{5} $ & $ 5.34\times10^{5} $ & $ 1.28 $ & $ 1.14\times10^{6} $ & M1 \\ \hline
M1S & $ 1.07\times10^{3} $ & $ 44 $ & $ 7.71\times10^{3} $ & $ 0.966 $ & $ 3.7\times10^{3} $ & $ 5.32\times10^{5} $ & $ 1.29 $ & $ 1.14\times10^{6} $ & M1 \\ \hline
M1T & $ 1.34\times10^{3} $ & $ 21 $ & $ 3.17\times10^{3} $ & $ 0.948 $ & $ 1.36\times10^{3} $ & $ 5.23\times10^{5} $ & $ 1.36 $ & $ 1.14\times10^{6} $ & M1 \\ \hline
M1U & $ 1.6\times10^{3} $ & $ 24 $ & $ 6.6\times10^{3} $ & $ 0.871 $ & $ 5.93\times10^{5} $ & $ 5.11\times10^{5} $ & $ 1.45 $ & $ 1.14\times10^{6} $ & M1 \\ \hline
M1V & $ 1.63\times10^{3} $ & $ 9 $ & $ 4.99\times10^{3} $ & $ 0.795 $ & $ 2.14\times10^{6} $ & $ 4.09\times10^{5} $ & $ 2.54 $ & $ 1.16\times10^{6} $ & M1 \\ \hline
M1W & $ 1.58\times10^{3} $ & $ 13 $ & $ 3.81\times10^{3} $ & $ 0.224 $ & $ 1.43\times10^{7} $ & $ 6.71\times10^{4} $ & $ 0.82 $ & $ 1.52\times10^{5} $ & M1 \\ \hline
M1X & $ 1.22\times10^{3} $ & $ 68 $ & $ 3.24\times10^{3} $ & $ 0.175 $ & $ 9.33\times10^{5} $ & $ 6.42\times10^{4} $ & $ 1.89 $ & $ 1.47\times10^{5} $ & M1 \\ \hline
M1Y & $ 1.26\times10^{3} $ & $ 33 $ & $ 1.26\times10^{3} $ & $ 0.066 $ & $ 4.87\times10^{4} $ & $ 5.66\times10^{4} $ & $ 2.75 $ & $ 1.44\times10^{5} $ & M1 \\ \hline
\end{longtable}

\tref{tab:rang} shows the radius at which the source resides in the cluster $r_\ast$, the angle between the velocity of the source and the line of sight between the source and the observer $\theta$, the inclination of the source $\delta$ corresponding to the angle between the line of sight and the angular momentum of the source, and the angle between the velocity of the source and its angular momentum $\nu$. All numbers were randomly generated assuming a square of uniform distribution for the radius with $r_\ast\in[0,1)$ and uniform distributions for the angle with $\theta\in[0,\pi)$, $\delta\in[\pi/36,35\pi/36)$, and $\nu\in[0,\pi)$. The interval of $\delta$ is set so that the ``high-inclination'' assumption for the aberrational phase shift is fulfilled~\citep{torres-orjuela_chen_2020} and we check that $\theta$, $\delta$, and $\nu$ fulfill the inequality $(\cos(\theta)-\cos(\delta)\cos(\nu))^2\leq\sin^2(\nu)\sin^2(\delta)$ to have a non-negative value in the square root in \eq{eq:abe}.

\begin{longtable}{| p{0.8cm} || p{1.55cm} | p{1.55cm} | p{1.55cm} | p{1.55cm} ||| p{0.8cm} || p{1.55cm} | p{1.55cm} | p{1.55cm} | p{1.55cm} |}
\caption{A list of the radius at which the source resides in the cluster $r_\ast$, the angle between the velocity of the source and the line of sight $\theta$, the inclination of the source $\delta$, and the angle between the velocity of the source and its angular momentum $\nu$.}\label{tab:rang} \\
\hline
ID & $r_\ast\,{\rm [R_h]}$ & $\theta\,{\rm [\pi]}$ & $\delta\,[\pi]$ & $\nu\,{\rm [\pi]}$ & ID & $r_\ast\,{\rm [R_h]}$ & $\theta\,{\rm [\pi]}$ & $\delta\,[\pi]$ & $\nu\,{\rm [\pi]}$ \\ \hline 
MWA & $ 1.52\times10^{-1} $ & $ 2.78\times10^{-1} $ & $ 5.29\times10^{-1} $ & $ 5.26\times10^{-1} $ & AGA & $ 2.8\times10^{-1} $ & $ 1.82\times10^{-1} $ & $ 2.78\times10^{-1} $ & $ 3.14\times10^{-1} $ \\ \hline
MWB & $ 4.51\times10^{-2} $ & $ 3.2\times10^{-1} $ & $ 4.26\times10^{-1} $ & $ 2.14\times10^{-1} $ & AGB & $ 3.66\times10^{-1} $ & $ 2.81\times10^{-1} $ & $ 5.12\times10^{-1} $ & $ 3.65\times10^{-1} $ \\ \hline
MWC & $ 3.94\times10^{-1} $ & $ 2.49\times10^{-1} $ & $ 2.11\times10^{-1} $ & $ 3.52\times10^{-1} $ & AGC & $ 1.6\times10^{-5} $ & $ 2.52\times10^{-1} $ & $ 4.38\times10^{-1} $ & $ 6.12\times10^{-1} $ \\ \hline
MWD & $ 7.01\times10^{-1} $ & $ 6.12\times10^{-1} $ & $ 4.51\times10^{-1} $ & $ 8.35\times10^{-1} $ & AGD & $ 1.48\times10^{-1} $ & $ 7.27\times10^{-1} $ & $ 5.85\times10^{-1} $ & $ 4.3\times10^{-1} $ \\ \hline
MWE & $ 2.21\times10^{-1} $ & $ 6.5\times10^{-2} $ & $ 4.61\times10^{-1} $ & $ 4.35\times10^{-1} $ & AGE & $ 8.79\times10^{-2} $ & $ 1.67\times10^{-1} $ & $ 7.02\times10^{-2} $ & $ 1.66\times10^{-1} $ \\ \hline
MWF & $ 4.87\times10^{-1} $ & $ 6.75\times10^{-1} $ & $ 8.85\times10^{-1} $ & $ 3.93\times10^{-1} $ & AGF & $ 2.18\times10^{-1} $ & $ 3.35\times10^{-1} $ & $ 3.09\times10^{-1} $ & $ 3.5\times10^{-1} $ \\ \hline
MWG & $ 1.62\times10^{-1} $ & $ 4.53\times10^{-1} $ & $ 3.38\times10^{-1} $ & $ 6.77\times10^{-1} $ & AGG & $ 1.45\times10^{-3} $ & $ 1.4\times10^{-1} $ & $ 7.11\times10^{-1} $ & $ 7.11\times10^{-1} $ \\ \hline
MWH & $ 6.22\times10^{-2} $ & $ 2.37\times10^{-1} $ & $ 7.34\times10^{-1} $ & $ 9.43\times10^{-1} $ & AGH & $ 1.3\times10^{-1} $ & $ 3.55\times10^{-1} $ & $ 4.49\times10^{-1} $ & $ 2.95\times10^{-1} $ \\ \hline
MWI & $ 5.37\times10^{-1} $ & $ 2.3\times10^{-1} $ & $ 1.38\times10^{-1} $ & $ 1.99\times10^{-1} $ & AGI & $ 5.58\times10^{-2} $ & $ 5.04\times10^{-1} $ & $ 1.38\times10^{-1} $ & $ 5.33\times10^{-1} $ \\ \hline
MWJ & $ 2.58\times10^{-1} $ & $ 2.47\times10^{-1} $ & $ 6.82\times10^{-1} $ & $ 5.94\times10^{-1} $ & AGJ & $ 7.51\times10^{-1} $ & $ 3.09\times10^{-1} $ & $ 6.58\times10^{-1} $ & $ 5.71\times10^{-1} $ \\ \hline
MWK & $ 3.97\times10^{-2} $ & $ 2.66\times10^{-1} $ & $ 2.72\times10^{-1} $ & $ 1.71\times10^{-1} $ & AGK & $ 6.46\times10^{-2} $ & $ 6.42\times10^{-1} $ & $ 7.27\times10^{-1} $ & $ 1.66\times10^{-1} $ \\ \hline
MWL & $ 2.22\times10^{-1} $ & $ 5.2\times10^{-1} $ & $ 9.63\times10^{-1} $ & $ 4.51\times10^{-1} $ & AGL & $ 1.96\times10^{-3} $ & $ 2.72\times10^{-1} $ & $ 5.09\times10^{-1} $ & $ 6.29\times10^{-1} $ \\ \hline
MWM & $ 4.4\times10^{-2} $ & $ 3.93\times10^{-1} $ & $ 2.39\times10^{-1} $ & $ 1.94\times10^{-1} $ & AGM & $ 3.54\times10^{-2} $ & $ 4.67\times10^{-1} $ & $ 4.41\times10^{-1} $ & $ 7.41\times10^{-1} $ \\ \hline
MWN & $ 1.05\times10^{-1} $ & $ 6.33\times10^{-1} $ & $ 8.66\times10^{-1} $ & $ 2.96\times10^{-1} $ & AGN & $ 2.3\times10^{-1} $ & $ 5.19\times10^{-2} $ & $ 4.81\times10^{-1} $ & $ 4.89\times10^{-1} $ \\ \hline
MWO & $ 6.53\times10^{-1} $ & $ 5.52\times10^{-1} $ & $ 9.37\times10^{-1} $ & $ 4.58\times10^{-1} $ & AGO & $ 8.38\times10^{-1} $ & $ 1.68\times10^{-1} $ & $ 3.76\times10^{-1} $ & $ 4.74\times10^{-1} $ \\ \hline
MWP & $ 1.77\times10^{-1} $ & $ 1.68\times10^{-1} $ & $ 8.5\times10^{-1} $ & $ 8.74\times10^{-1} $ & AGP & $ 7.27\times10^{-2} $ & $ 8\times10^{-1} $ & $ 9.14\times10^{-1} $ & $ 1.22\times10^{-1} $ \\ \hline
MWQ & $ 8.75\times10^{-2} $ & $ 3.84\times10^{-1} $ & $ 3.2\times10^{-1} $ & $ 3.63\times10^{-1} $ & AGQ & $ 2.87\times10^{-1} $ & $ 7.97\times10^{-1} $ & $ 2.24\times10^{-1} $ & $ 9.67\times10^{-1} $ \\ \hline
MWR & $ 1.7\times10^{-1} $ & $ 3.85\times10^{-1} $ & $ 3.16\times10^{-1} $ & $ 2.41\times10^{-1} $ & AGR & $ 1.2\times10^{-1} $ & $ 4.53\times10^{-1} $ & $ 5.29\times10^{-1} $ & $ 5.84\times10^{-1} $ \\ \hline
MWS & $ 3.02\times10^{-1} $ & $ 3.93\times10^{-1} $ & $ 3.74\times10^{-1} $ & $ 8.69\times10^{-2} $ & AGS & $ 8.04\times10^{-3} $ & $ 3.56\times10^{-1} $ & $ 6.35\times10^{-1} $ & $ 9.3\times10^{-1} $ \\ \hline
MWT & $ 7.48\times10^{-1} $ & $ 2.91\times10^{-1} $ & $ 7.24\times10^{-1} $ & $ 7.91\times10^{-1} $ & AGT & $ 4.55\times10^{-2} $ & $ 8.9\times10^{-1} $ & $ 7.77\times10^{-1} $ & $ 1.69\times10^{-1} $ \\ \hline
MWU & $ 6.72\times10^{-2} $ & $ 7.39\times10^{-1} $ & $ 7.53\times10^{-1} $ & $ 4.56\times10^{-2} $ & AGU & $ 4.27\times10^{-1} $ & $ 3.31\times10^{-1} $ & $ 4.57\times10^{-1} $ & $ 7.75\times10^{-1} $ \\ \hline
MWV & $ 1.36\times10^{-1} $ & $ 3.43\times10^{-1} $ & $ 3.87\times10^{-1} $ & $ 7.29\times10^{-1} $ & AGV & $ 2.49\times10^{-1} $ & $ 1.31\times10^{-1} $ & $ 2.7\times10^{-1} $ & $ 3.87\times10^{-1} $ \\ \hline
MWW & $ 3.13\times10^{-1} $ & $ 2.66\times10^{-1} $ & $ 1.91\times10^{-1} $ & $ 2.06\times10^{-1} $ & AGW & $ 6.11\times10^{-1} $ & $ 7.14\times10^{-1} $ & $ 6.79\times10^{-1} $ & $ 6.77\times10^{-2} $ \\ \hline
MWX & $ 3.42\times10^{-3} $ & $ 4.04\times10^{-1} $ & $ 3.21\times10^{-1} $ & $ 6.97\times10^{-1} $ & AGX & $ 1.78\times10^{-1} $ & $ 4.37\times10^{-1} $ & $ 5.56\times10^{-1} $ & $ 6.2\times10^{-1} $ \\ \hline
MWY & $ 3.8\times10^{-1} $ & $ 5.44\times10^{-1} $ & $ 3.28\times10^{-1} $ & $ 7.93\times10^{-1} $ & AGY & $ 2.95\times10^{-1} $ & $ 2.27\times10^{-1} $ & $ 5.46\times10^{-1} $ & $ 4.36\times10^{-1} $ \\ \hline
MCA & $ 8.44\times10^{-1} $ & $ 1.62\times10^{-1} $ & $ 5.8\times10^{-1} $ & $ 6.98\times10^{-1} $ & M1A & $ 2.24\times10^{-1} $ & $ 4.66\times10^{-1} $ & $ 5.09\times10^{-1} $ & $ 3.18\times10^{-1} $ \\ \hline
MCB & $ 2.11\times10^{-1} $ & $ 1.69\times10^{-1} $ & $ 5.99\times10^{-1} $ & $ 7.22\times10^{-1} $ & M1B & $ 3.18\times10^{-1} $ & $ 7.69\times10^{-1} $ & $ 7.6\times10^{-1} $ & $ 3.87\times10^{-1} $ \\ \hline
MCC & $ 8.82\times10^{-2} $ & $ 3.29\times10^{-1} $ & $ 4.93\times10^{-1} $ & $ 4.96\times10^{-1} $ & M1C & $ 2.08\times10^{-1} $ & $ 5.86\times10^{-1} $ & $ 5.45\times10^{-1} $ & $ 8.45\times10^{-1} $ \\ \hline
MCD & $ 5.61\times10^{-1} $ & $ 4.7\times10^{-1} $ & $ 6.73\times10^{-1} $ & $ 2.57\times10^{-1} $ & M1D & $ 1.69\times10^{-1} $ & $ 2.41\times10^{-1} $ & $ 3.26\times10^{-1} $ & $ 1.92\times10^{-1} $ \\ \hline
MCE & $ 2.72\times10^{-1} $ & $ 6.39\times10^{-1} $ & $ 4.53\times10^{-1} $ & $ 2.72\times10^{-1} $ & M1E & $ 5.18\times10^{-1} $ & $ 5.5\times10^{-1} $ & $ 7.32\times10^{-1} $ & $ 3.41\times10^{-1} $ \\ \hline
MCF & $ 4.48\times10^{-1} $ & $ 4.67\times10^{-1} $ & $ 6.19\times10^{-1} $ & $ 7.99\times10^{-1} $ & M1F & $ 7.68\times10^{-1} $ & $ 2.29\times10^{-1} $ & $ 3.16\times10^{-1} $ & $ 4.82\times10^{-1} $ \\ \hline
MCG & $ 2.52\times10^{-1} $ & $ 1.42\times10^{-1} $ & $ 1.87\times10^{-1} $ & $ 2.07\times10^{-1} $ & M1G & $ 3.01\times10^{-1} $ & $ 4.63\times10^{-1} $ & $ 2.37\times10^{-1} $ & $ 6.06\times10^{-1} $ \\ \hline
MCH & $ 2.8\times10^{-2} $ & $ 2.57\times10^{-1} $ & $ 3.56\times10^{-1} $ & $ 5.31\times10^{-1} $ & M1H & $ 2.54\times10^{-1} $ & $ 5.52\times10^{-1} $ & $ 4.86\times10^{-1} $ & $ 4.95\times10^{-1} $ \\ \hline
MCI & $ 1.26\times10^{-1} $ & $ 1.07\times10^{-1} $ & $ 2.05\times10^{-1} $ & $ 2.34\times10^{-1} $ & M1I & $ 1.79\times10^{-1} $ & $ 7.84\times10^{-1} $ & $ 3.76\times10^{-1} $ & $ 7.63\times10^{-1} $ \\ \hline
MCJ & $ 3\times10^{-1} $ & $ 9.08\times10^{-1} $ & $ 4.54\times10^{-1} $ & $ 4.71\times10^{-1} $ & M1J & $ 1.35\times10^{-1} $ & $ 4.03\times10^{-1} $ & $ 5.18\times10^{-1} $ & $ 1.32\times10^{-1} $ \\ \hline
MCK & $ 1.48\times10^{-1} $ & $ 4.91\times10^{-1} $ & $ 5.18\times10^{-1} $ & $ 4.34\times10^{-1} $ & M1K & $ 1.47\times10^{-2} $ & $ 3.85\times10^{-1} $ & $ 2.4\times10^{-1} $ & $ 6.04\times10^{-1} $ \\ \hline
MCL & $ 2.17\times10^{-1} $ & $ 6.08\times10^{-1} $ & $ 7.03\times10^{-1} $ & $ 5.36\times10^{-1} $ & M1L & $ 3.21\times10^{-2} $ & $ 7.6\times10^{-1} $ & $ 7.86\times10^{-1} $ & $ 3.83\times10^{-1} $ \\ \hline
MCM & $ 1.21\times10^{-1} $ & $ 7.4\times10^{-2} $ & $ 5.02\times10^{-1} $ & $ 4.78\times10^{-1} $ & M1M & $ 1.21\times10^{-1} $ & $ 8.32\times10^{-1} $ & $ 5.23\times10^{-1} $ & $ 4.4\times10^{-1} $ \\ \hline
MCN & $ 3.48\times10^{-1} $ & $ 3.1\times10^{-1} $ & $ 4.74\times10^{-1} $ & $ 7.05\times10^{-1} $ & M1N & $ 3.57\times10^{-3} $ & $ 3.57\times10^{-1} $ & $ 4.12\times10^{-1} $ & $ 8.22\times10^{-2} $ \\ \hline
MCO & $ 5.1\times10^{-1} $ & $ 7.22\times10^{-1} $ & $ 8.21\times10^{-1} $ & $ 4.02\times10^{-1} $ & M1O & $ 3.85\times10^{-2} $ & $ 2.85\times10^{-2} $ & $ 3.63\times10^{-1} $ & $ 3.88\times10^{-1} $ \\ \hline
MCP & $ 2.56\times10^{-1} $ & $ 5.08\times10^{-1} $ & $ 6.05\times10^{-1} $ & $ 7.06\times10^{-1} $ & M1P & $ 9.17\times10^{-3} $ & $ 5.57\times10^{-1} $ & $ 6.57\times10^{-1} $ & $ 4.58\times10^{-1} $ \\ \hline
MCQ & $ 2.12\times10^{-1} $ & $ 4.42\times10^{-1} $ & $ 3.49\times10^{-1} $ & $ 7.13\times10^{-1} $ & M1Q & $ 3.39\times10^{-1} $ & $ 2.15\times10^{-1} $ & $ 1.57\times10^{-1} $ & $ 3.72\times10^{-1} $ \\ \hline
MCR & $ 4.52\times10^{-1} $ & $ 8.66\times10^{-1} $ & $ 4.34\times10^{-1} $ & $ 5.07\times10^{-1} $ & M1R & $ 2.63\times10^{-1} $ & $ 9.11\times10^{-2} $ & $ 3.85\times10^{-1} $ & $ 4.48\times10^{-1} $ \\ \hline
MCS & $ 6.04\times10^{-1} $ & $ 3.13\times10^{-1} $ & $ 5.57\times10^{-1} $ & $ 7.71\times10^{-1} $ & M1S & $ 3.33\times10^{-1} $ & $ 5.41\times10^{-1} $ & $ 3.6\times10^{-1} $ & $ 7.41\times10^{-1} $ \\ \hline
MCT & $ 4.21\times10^{-1} $ & $ 6.34\times10^{-1} $ & $ 4.58\times10^{-1} $ & $ 7.98\times10^{-1} $ & M1T & $ 2.44\times10^{-2} $ & $ 5.62\times10^{-1} $ & $ 1.85\times10^{-1} $ & $ 4.37\times10^{-1} $ \\ \hline
MCU & $ 9.16\times10^{-3} $ & $ 3.52\times10^{-1} $ & $ 7.86\times10^{-1} $ & $ 6.99\times10^{-1} $ & M1U & $ 6.24\times10^{-2} $ & $ 5.76\times10^{-1} $ & $ 8.89\times10^{-1} $ & $ 3.97\times10^{-1} $ \\ \hline
MCV & $ 3.28\times10^{-1} $ & $ 2.84\times10^{-1} $ & $ 4.44\times10^{-1} $ & $ 3.48\times10^{-1} $ & M1V & $ 4.26\times10^{-2} $ & $ 7.36\times10^{-1} $ & $ 6.14\times10^{-1} $ & $ 6.42\times10^{-1} $ \\ \hline
MCW & $ 4.57\times10^{-2} $ & $ 8.33\times10^{-1} $ & $ 4.45\times10^{-1} $ & $ 5.93\times10^{-1} $ & M1W & $ 3.45\times10^{-1} $ & $ 8.52\times10^{-1} $ & $ 5.85\times10^{-1} $ & $ 4.43\times10^{-1} $ \\ \hline
MCX & $ 2.97\times10^{-1} $ & $ 4.52\times10^{-1} $ & $ 7.45\times10^{-1} $ & $ 7.79\times10^{-1} $ & M1X & $ 1.8\times10^{-2} $ & $ 7.96\times10^{-1} $ & $ 2.16\times10^{-1} $ & $ 9.36\times10^{-1} $ \\ \hline
MCY & $ 5.39\times10^{-2} $ & $ 1.63\times10^{-1} $ & $ 7.93\times10^{-1} $ & $ 8.79\times10^{-1} $ & M1Y & $ 4.09\times10^{-1} $ & $ 4.46\times10^{-1} $ & $ 5.7\times10^{-1} $ & $ 9.25\times10^{-1} $ \\ \hline
\end{longtable}

\tref{tab:det} shows the SNR $\rho$ and the match between the signal and a model not accounting for the IMRI's Brownian motion $\mathcal{M}$ for AION, LISA, and TianQin. The list only contains sources with an SNR of at least one in one of the detectors, and in cases where the SNR is zero, we do not specify the match.

\begin{longtable}{| p{1cm} || p{1.7cm} | p{1.7cm} | p{1.7cm} | p{1.7cm} | p{1.7cm} | p{1.7cm} |}
\caption{The SNR $\rho$ and the match $\mathcal{M}$ in AION, LISA, and TianQin for a subset of the IMRIs considered.}\label{tab:det} \\
\hline
ID & $\rho_{\rm AION}$ & $\rho_{\rm LISA}$ & $\rho_{\rm TianQin}$ & $\mathcal{M}_{\rm AION}$ & $\mathcal{M}_{\rm LISA}$ & $\mathcal{M}_{\rm TianQin}$ \\ \hline
MWI & $ 17.8 $ & $ 31.4 $ & $ 22.6 $ & $ 0.852 $ & $ 0.731 $ & $ 0.452 $ \\ \hline
MWL & $ 6.9 $ & $ 19.4 $ & $ 10.0 $ & $ 0.164 $ & $ 0.653 $ & $ 0.102 $ \\ \hline
MWR & $ 0.0 $ & $ 148.0 $ & $ 45.8 $ & $ 1.0 $ & $ 0.005 $ & $ 0.052 $ \\ \hline
MCF & $ 0.0 $ & $ 1.6 $ & $ 0.0 $ & $ 1.0 $ & $ 0.833 $ & $ 1.0 $ \\ \hline
MCH & $ 16.7 $ & $ 37.2 $ & $ 38.4 $ & $ 1.0 $ & $ 0.941 $ & $ 0.999 $ \\ \hline
MCO & $ 2.1 $ & $ 1.2 $ & $ 0.5 $ & $ 0.623 $ & $ 0.183 $ & $ 0.362 $ \\ \hline
AGB & $ 28.3 $ & $ 93.6 $ & $ 14.7 $ & $ 0.492 $ & $ 0.05 $ & $ 0.466 $ \\ \hline
AGW & $ 0.2 $ & $ 2.4 $ & $ 0.4 $ & $ 0.254 $ & $ 0.362 $ & $ 0.062 $ \\ \hline
M1B & $ 2.4 $ & $ 3.2 $ & $ 2.4 $ & $ 0.423 $ & $ 0.455 $ & $ 0.406 $ \\ \hline
\end{longtable}


\bibliography{alebib}{}
\bibliographystyle{aasjournal}



\end{document}